\documentclass[onecolumn,english,showpacs,preprintnumbers,amsmath,amssymb,floatfix,12pt]{revtex4}

\usepackage[T1]{fontenc}
\usepackage[latin9]{inputenc}
\usepackage{color}
\usepackage{array}
\usepackage{amstext}
\usepackage{graphicx}
\usepackage{esint}
\usepackage{rotating}
\usepackage{amsmath}
\usepackage{slashed}
\usepackage[font=small,labelfont=bf]{caption}
\usepackage[colorlinks = true,
linkcolor = blue,
urlcolor  = blue,
citecolor = blue,
anchorcolor = blue]{hyperref}

\makeatletter

\@ifundefined{textcolor}{}
{%
 \definecolor{BLACK}{gray}{0}
 \definecolor{WHITE}{gray}{1}
 \definecolor{RED}{rgb}{1,0,0}
 \definecolor{GREEN}{rgb}{0,1,0}
 \definecolor{BLUE}{rgb}{0,0,1}
 \definecolor{CYAN}{cmyk}{1,0,0,0}
 \definecolor{MAGENTA}{cmyk}{0,1,0,0}
 \definecolor{YELLOW}{cmyk}{0,0,1,0}
 }

\@ifundefined{definecolor}
 {\usepackage{color}}{}
\@ifundefined{definecolor}
 {\usepackage{color}}{}
\makeatother
\usepackage{babel}

\begin{document}


\title{Structure and tidal deformability of a hybrid star within the framework of the field correlator method  }

\author{S. Khanmohamadi}
\email{s.khanmohamadi@ut.ac.ir}
\author{H. R. Moshfegh}%
\email{hmoshfegh@ut.ac.ir}
\affiliation{%
	Department of Physics, University of Tehran, P.O.Box 14395-547, Tehran, Iran
}%
\author {S.~Atashbar Tehrani}
\email{Atashbar@ipm.ir}
\affiliation {
	School of Particles and Accelerators, Institute for Research in Fundamental Sciences (IPM), P.O.Box
	19395-5531, Tehran, Iran}
\date{\today}

%
%
\begin{abstract}\label{abstract}
The structure of hybrid stars within the nonperturbative framework of the field correlator method, extended to zero-temperature limit as a quark model, has been studied. For the hadronic sector, we have used the lowest-order constraint variational method by employing $ AV_{18} $ two-body nucleon-nucleon interaction supplemented by the phenomenological Urbana-type three-body force. 
For an adapted value of  the gluon condensate, $ G_{2}=0.006 $ GeV$ ^{4} $, which gives the critical temperature of about $ T_{c}\sim170 $ MeV, stable hybrid stars with a maximum mass of $ 2.04M_{\odot}  $ are predicted. The stability of hybrid star has been investigated for a wide range of gluon condensate value, $ G_{2} $, and quark-antiquark potential, $V_{1}$. A hybrid equation of state fulfills the constraints on tidal deformability and hence on the radii of the stars, extracted from the binary GW170817.  Moreover, tidal deformability for different chirp masses and different binary mass ratios of hybrid stars have been studied. The mass-radius relation satisfies the new constraint obtained from the neutron star interior composition explorer (NICER).
 A comprehensive analysis on the structure of a hybrid star and also its compactness, tidal Love number, and tidal deformability has been conducted for  several parameter sets of the quark equation of state. The influence of different crustal equations of state on the mentioned quantities has been studied. Our calculations suggest the value of quark-antiquark potential, $ V_{1} $, to be around 0.08 GeV. 
The results achieved in this study are in strong concurrence with the other calculations reported on this subject.

\end{abstract}
%


\maketitle


%
\section{Introduction}\label{Introduction}
In the last few decades, a great effort has been made to understand the properties of nuclear matter at densities higher than nuclear densities. From heavy ion collisions and astrophysical observations of compact objects, many attempts have been made to determine the equation of state (EOS) of dense nuclear matter in both hadronic and quark phases.
The probable appearance of quark degrees of freedom in the interior of heavy neutron stars (NSs) is one of the most debated issues in the context of the compact stars  ~\cite{Glendenning:2001pe,Witten:1984rs,Baym:1985tn}. By the discovery of two massive NSs~\cite{Demorest:2010bx,Antoniadis:2013pzd,Lynch:2012vv,vanKerkwijk:2010mt,Fonseca:2016tux}, the question of whether quark matter exists in the core of neutron stars has newly received interest~\cite{Hempel:2009vp,Burgio:2002sn,Burgio:2015zka,Weber:2004kj,Blaschke:2018mqw,Radzhabov:2010dd,Blaschke:2007ri}.

The study of properties of NSs concerns the high-density and low-temperature region of the phase diagram, and in particular, it requires the QCD nonperturbative EOS at low temperature and large chemical potential, where the essential theoretical lattice formalism of QCD is inapplicable. Due to the lack of lattice data, analytic approaches such as the MIT bag model~\cite{Farhi:1984qu,Alcock:1986hz} and the Nambu-Jona-Lasinio (NJL) model~\cite{Buballa:2003qv} are mostly used in the high-density regions. 

The MIT bag model  provides a mechanism for natural confinement by the inclusion of phenomenological confining pressure, which is the difference in energy density between the peturbative vacuum and true vacuum, named the bag constant, $ B $. The NJL model contains one of the basic symmetries of QCD, namely chiral symmetry. The most important feature of this model is its nontrivial vacuum in which the chiral symmetry is broken dynamically by the spontaneous mass generation. The NJL model is applicable in vacuum as well as in high densities but not in the hadronic phase in between, because of the lack of confinement due to the lack of gluon degrees of freedom in this model. 

 In our previous study~\cite{Khanmohamadi:2019jky}, we investigated the properties of hybrid stars (HSs) within the Nambu-Jona-Lasinio (NJL) and MIT bag models. Within both quark models, stable HSs with pure quark cores were predicted-however, with a the maximum mass lower than $ 2M_{\odot} $. HSs with the maximum mass compatible with the observations were predicted, although, they were found to be unstable. Since the quark phase is unable to support the increasing central pressure due to gravity, the instability is manifested by a cusp at the maximum mass of the mass-radius relation~\cite{Baldo:2002ju}. However, the radii and tidal deformability of HSs were in the same range as deduced from the gravitational wave data of the binary GW170817.
 
 The general feature of many quark models, which is their serious drawback, is their inability to give predictions for the full temperature-chemical potential range. One of the few exceptions is the field correlator method (FCM)~\cite{DiGiacomo:2000irz,Simonov:2007jb,Baldo:2008en}, which in principle could cover the full phase diagram panel. Besides, the same method includes from first principles the property of confinement (in contrast to the NJL model), which seems to have a role in the stability of the predicted HS~\cite{Baldo:2002ju,Baldo:2006bt}.

Microscopic theories of baryonic matter have been developed in the last few decades in contrary to quark matter case. The lowest-order constraint variational method (LOCV), which is a pure variational technique in the study of the bulk properties of many fermion systems~\cite{Owen:1975xh,Modarres:1979jk,Reid:1968sq,Green:1978fj}, is employed as the nucleonic model of our study. This approach is extended in such a way to enable one to calculate the properties of asymmetric nuclear matter, neutron matter and beta stable matter EOSs at both zero and finite temperatures by using more sophisticated potentials ~\cite{Moshfegh:2005rom,Moshfegh:2007mxh,Modarres:2000nk,Modarres:2002ns}. Besides this, the thermodynamic properties of nuclear matter at both zero and finite temperatures are calculated by considering relativistic corrections in this formalism~\cite{Zaryouni:2010p,Zaryouni:2014fsa}. It is well known that the bare two-body nucleon-nucleon (2BF) interactions cannot reproduce the saturation properties of nuclear matter. The LOCV method is capable dealing with the three-body forces as well~\cite{Goudarzi:2015dax,Goudarzi:2016uos}. Recently we have shown that by employing the Urbana-type (UIX) three-body forces (TBF), one can obtain the correct saturation quantities such as binding energy, saturation density and symmetry properties like $ E_{sym}(\rho_{0},L,$ and $ K_{sym}) $. Also, the NSs with masses above $ 2M_{\odot} $ are predicted within the LOCV formalism employing $ AV_{18} $ supplemented by TBF in Urbana-type ~\cite{Goudarzi:2016uos} and chiral symmetry~\cite{Goudarzi:2019orb}. The EOS of Hypernuclear matter is produced within LOCV method~\cite{Shahrbaf:2019bef,Shahrbaf:2019wex}. Newly, the HS structure is studied within the framework of the LOCV method combined with the three-flavor version of the NJL model for several parameter sets of this model~\cite{Khanmohamadi:2019jky}. Moreover, the phase transition of hypernuclear matter to the two-flavor version of the nonlocal NJL model is recently being studied~\cite{Shahrbaf:2019vtf}. 
 
 In the study of HS structure, the nucleon-quark phase transition plays an important role. We restrict our study to analyze sharp  hadron-to-quark matter phase transition.
 It may happen that a hadron-quark mixed phase is unlikely to be stable for a reasonable value of the surface tension; this situation is closer to the Maxwell construction case~\cite{Maruyama:2006jk,Alford:2001zr,Neumann:2002jm}, where two pure phases are in direct contact with each other, and it shows a sharp phase transition behavior.  
 
 In this paper we have employed the nonperturbative EOS of quark-gluon plasma which was derived in the framework of the FCM for describing quark matter~\cite{DiGiacomo:2000irz,Simonov:2007jb}. The FCM is a nonpertubative approach which provides a natural treatment of the dynamics of confinement and transition to deconfinement phase in terms of color electric and color magnetic correlators. The quark-antiquark ($ q\bar{q} $) potential, $ V_{1} $, and the gluon condensate, $ G_{2} $, are the parameters of the model whose numerical values are partially supported by lattice simulations at small chemical potentials ~\cite{Simonov:2007jb,Doring:2005ih} and the QCD sum rules~\cite{Shifman:1978by}, respectively.
 
 In recent years, besides the maximum-mass constraints on compact stars' EOSs,  $ 2.01^{+0.04}_{-0.04}\leq{M}_{TOV}/{M\odot}\lesssim 2.16^{+0.17}_{-0.15}$, there exists a new constraint on tidal deformability, and hence on the radii of compact stars set by the binary NS system GW170817~\cite{Most:2018hfd} . With the first direct detection of both gravitational and electromagnetic radiation from the binary NS merger GW170817 on August 17 2017 (recorded by the Advanced LIGO and Virgo network of gravitational-wave recorders~\cite{TheLIGOScientific:2017qsa,Abbott:2018wiz,Abbott:2018exr}), we are facing a new important feature of astronomy which could help us to understand the origin of these phenomena~\cite{Negele:1971vb,Drischler:2016djf,Annala:2017llu,Drischler:2017wtt}. By applying the tidal deformability constraints on the EOS, GW170817 provides a new essential insight to understand the physics of matter under extreme density conditions. The influence of the perturbing tidal field of the companion of a NS is reflected in the tidal deformability, $ \Lambda $. These new constraints will be investigated on the several hybrid star EOSs in this paper, and we will go further to present the predictions for tidal deformability with different chirp masses and different binary mass ratios. 
  We will also check the new constraint on the mass-radius relation extracted from the neutron star interior composition explorer (NICER). The effect of different crustal EOSs on the structure and tidal deformability of a HS will be studied too.

 The paper is organized as follows: In Sec.~\ref{II} we briefly describe the EOS of the nucleonic sector in beta equilibrium at zero temperature within the LOCV method. Section.~\ref{III} is devoted to  the quark matter EOSs according to the FCM. In Sec.~\ref{IV-A}, by using these models, the EOS of a HS is proposed assuming the Maxwell construction. The structure of the HS is presented in Sec.~\ref{IV-B}, and in Sec.~\ref{IV-C} we are concerned with the calculation of the tidal deformability, new constraints on the mass-radius relation extracted from NICER and the effect of different crustal EOSs on the structure of the hybrid stars. The summary and concluding remarks are presented in Sec.~\ref{V}.

\section{ Hadronic phase: LOCV approach}\label{II}

 The LOCV model is a microscopic model based on cluster expansion and is in good agreement with empirical nuclear saturation properties. 

The first step in the LOCV formalism is to consider a trial wave function for the $ N $-body interacting system at zero temperature. Such trial wave function is given by
\begin{equation}
\Psi(1\ldots N)=F(1\ldots N)\Phi(1\ldots N),
\end{equation}
where $\Phi(1\ldots N) $ is a noninteracting ground-state wave function of $ N $ independent nucleons and $F(1\ldots N)$ is an $ N $-body correlation operator considered in the Jastrow approximation.

In general, the nuclear Hamiltonian is read as sum of the nonrelativistic single-particle kinetic energy and the nucleonic potential
\begin{equation}
H=\sum _{i}\frac{p_{i} ^{2} }{2m_{i}}  +\sum _{i<j}V(ij)+\sum _{i<j<k}V(ijk)+....
\end{equation}
So the baryonic energy expectation value $ E_{B} $ can be written as
\begin{equation}\label{E}
E_{B}[f]=\frac{1}{N} \frac{{\left\langle \Psi  \right|} H{\left| \Psi  \right\rangle} }{{\left\langle \Psi  \mathrel{\left| \vphantom{\Psi  \Psi }\right.\kern-\nulldelimiterspace} \Psi  \right\rangle} } =E_{1} +E_{MB} \cong E_{1} +E_{2},
\end{equation}
in which $E_{1}$ is the one-body energy and $E_{2}$ is 
 the two-body energy. Higher-order terms in the cluster expansion series are negligible~\cite{Moshfegh:2005rom}. $ E_{2} $ is minimized with respect to the channel correlation functions but subjected to the normalization constraint~\cite{Moshfegh:2005rom,Modarres,Owen:1977uun}, which introduces the Lagrange multipliers in the formalism. The procedure of minimizing $ E_{2} $ provides a number of Euler-Lagrange differential  equations for two-body correlation operators. Solving these equations leads to the determination of correlation functions and the two-body cluster energy. By the inclusion of TBF in the nuclear Hamiltonian, the problem of incorrectly reproducing the saturation properties of cold symmetric nuclear matter is resolved. In order to avoid the full three-body problem, the TBF (semiphenomenological UIX interaction ) is included via an effective two-body potential derived after averaging out the third particle, which is weighted by the LOCV two-body correlation functions at a given baryonic density $\rho_{B}$. For more details, see Refs. \cite{Goudarzi:2015dax,Goudarzi:2016uos}.

 The beta equilibrium condition should be imposed on the EOS of the NS, since the density of the system is high enough for nuclei to dissolve to form an interacting system of nucleons and leptons. As the system survives longer than the timescale of weak interactions, it reaches equilibrium with respect to the $ \beta $ decay $ n=p+e+\nu\bar{}_{e} $ and its inverse reaction.
 
 By solving the $ \beta $ equilibrium conditions,
$ \mu _{n}=\mu _{p} +\mu _{e} $ and $ \mu _{e}=\mu _{\mu} $,
along with the charge neutrality condition $ \rho _{p}=\rho _{\mu}+\rho _{e} $ at any given baryon density,$ \rho_{B} $, the energy of $\beta $-stable matter, $ E $, written as the sum of the baryonic energy and leptonic energy can be determined (details of calculations can be found in related references of LOCV formalism). Leptons are supposed to be highly relativistic noninteracting particles.
The pressure of the NS matter as a function of baryonic density is calculated by using the following thermodynamic relation

\begin{equation}
P=\rho_{B}^{2}(\dfrac{\partial (E/N)}{\partial \rho_{B}})~.
\end{equation}

\section{Quark matter: FCM method }\label{III}
 In this section we briefly address the nonperturbative framework of the field correlator method. The FCM is a systematic method of computing non perturbative effects in the quenched approximation from some fundamental nonperturbative input. A set of field strength  correlators, namely\cite{DiGiacomo:2000irz},
\begin{eqnarray}\label{delta}
\Delta_{\mu_{1}\nu_{1},...,\mu_{n}\nu_{n}} =&&Tr<F_{\mu_{1}\nu_{1}}(x_{1})\Phi(x_{1},x_{2})F_{\mu_{2}\nu_{2}}(x_{2})...
\nonumber\\
&&~~~~~~~~F_{\mu_{n}\nu_{n}} (x_{n}) \Phi(x_{n},x_{1})>
\end{eqnarray} 
 is chosen as the nonperturbative input, where $ F_{\mu\nu} $'s are the field strength tensors and  $ \Phi(x,y) $'s are the phase factors, introduced for the gauge invariance condition. The main idea, which is proposed in refs.~\cite{Dosch:1987sk,Dosch:1988ha,Simonov:1987rn}, is to use the gauge-invariant quantities in Eq.~(\ref{delta}) as a dynamical input in the nonperturbative domain and to describe gauge-invariant observables through Eq.~(\ref{delta}) via the cluster expansion. Moreover, a systematic cluster expansion can be performed, and the first term, named the Gaussian correlator, gives a good qualitative description of most nonperturbative phenomena, while higher cumulants can be considered as corrections. It is shown that these corrections are not large and contribute around a few percent of the total effects~\cite{DelDebbio:1994zn,DiGiacomo:1989yp,DiGiacomo:1990hc,Bali:1997aj}. Therefore, one obtains a theory with a simple but fundamental input - Gaussian approximation - and the corresponding formalism is called Gaussian dominance approximation or the Gaussian stochastic model of QCD vacuum. The method can be called "fundamental phenomenology," since it uses correlators (actually the lowest-order one) in Eq.~(\ref{delta}) as the dynamical input which is given by lattice measurements. The necessary nonperturbative information enters via string tension $ \sigma $, which is an integral characteristic of the Gaussian correlator, and as a result, one can define the hadron in terms of one parameter. 
 In the FCM method, the Euclidean vacuum picture of QCD fields is considered. The results of QCD sum rules and quarkonium spectrum analysis show that the gluon vacuum is dense~\cite{DiGiacomo:2000irz} and the value of the gluon condensate $G_2$ is, 
 
\begin{equation}\label{g2}
G_{2}\equiv\dfrac{\alpha_{s}}{\pi}<F_{\mu\nu}^{a}F_{\mu\nu}^{a}>~\sim0.012~\text{GeV}^{4}
\end{equation} 
with 50\% uncertainty. Each point of the phase diagram can be characterized by the values of condensates, describing the symmetry-breaking pattern. The simplest condensates are given by the nonperturbative gluon and quark condensates: $<\dfrac{\alpha_{s}}{\pi}F_{\mu\nu}F_{\mu\nu}>$ and $<\bar\Psi\Psi> $.

A dynamical characteristic of such stochastic vacuum is given by a set of gauge-invariant correlators, which in the non-Abelian case have the form
\begin{eqnarray}\label{deltanon}
\Delta_{1,2,...,n} =&&\dfrac{1}{N_{c}}<TrG_{\mu_{1}\nu_{1}}(x_{1},x_{0})G_{\mu_{2}\nu_{2}}(x_{2},x_{0})...
\nonumber\\
&&~~~~~~~~G_{\mu_{n}\nu_{n}} (x_{n},x_{0}) >
\end{eqnarray} 
where 
\begin{equation}\label{gmu}
G_{\mu_{k}\nu_{k}}(x_{k},x_{0}) =\Phi(x_{0},x_{k})F_{\mu_{k}\nu_{k}}(x_{k})\Phi(x_{k},x_{0})
\end{equation} 

and phase factors $ \Phi $ are defined as follows:

\begin{equation}\label{phi}
\Phi(x,y) =P~exp~i\int_{y}^{x} A_{\mu}dz_{\mu}
\end{equation} 
with the path integral taken along some curves, connecting the initial and the final points. Although, the functions $\Delta_{1,2,...,n}  $ depend on the form of the contour, this dependence has to be canceled in physical quantities~\cite{DiGiacomo:2000irz}.

The most attractive feature of the nonlocal averages~[Eq.\ref{deltanon}] is its gauge invariance as compared with the case of usual gauge field Green's functions $ <A(x)A(y)...A(z)> $.

The dynamics of confinement is described by Gaussian color electric [$ D^{E}(x),D_{1}^{E}(x) $] and color magnetic [$ D^{H}(x),D_{1}^{H}(x) $] gauge invariant field correlators. The main quantity which governs the nonperturbative dynamics of deconfinement is given by the two point functions:
\begin{eqnarray}\label{g}
&&g^{2}<Tr_{f}[E_{i}(x)\varPhi(x,y)E_{k}(y)\varPhi(y,x)]>
\nonumber\\
&&=\delta_{ik}[D^{E}+D_{1}^{E}+z_{4}^{2}\dfrac{\partial D_{1}^{E}}{\partial z_{4}^{2}}]+z_{i}z_{k}\dfrac{\partial D_{1}^{E}}{\partial \vec{u}^{2}}~,
\nonumber\\
&&g^{2}<Tr_{f}[H_{i}(x)\varPhi(x,y)H_{k}(y)\varPhi(y,x)]>
\nonumber\\
&&=\delta_{ik}[D^{H}+D_{1}^{H}+\vec{u}^{2}\dfrac{\partial D_{1}^{H}}{\partial \vec{u}^{2}}]-z_{i}z_{k}\dfrac{\partial D_{1}^{H}}{\partial \vec{u}^{2}}~,
\end{eqnarray}
where $ z=x-y $ 
, and
  
\begin{equation}\label{varphi}
\varPhi(x,y) =P~exp~ig\int_{y}^{x} A_{\mu}dz_{\mu}
\end{equation} 
is the parallel transporter to assuring gauge invariance.

In the confined phase (below $ T_{c} $), $ D^{E}(x) $ is responsible for the confinement with string tension $ \sigma^{E}=\frac{1}{2}\int D^{E}(x) d^{2}x$. In the deconfinment phase (above $ T_{c} $ ), $ D^{E}(x) $ vanishes while $ D_{1}^{E}(x) $ remains nonzero, being responsible [together with the magnetic part due to $ D^{H}(x) $ and $ D_{1}^{H}(x) $] for nonperturbative dynamics of the deconfined phase. 

In the lattice calculations, the nonperturbative part of  $ D_{1}^{E}(x) $ is parameterized as follows~\cite{DiGiacomo:2000irz} 

\begin{equation}\label{D1Ex}
 D_{1}^{E}(x) =D_{1}^{E}(0)e^{-\lvert x \rvert/\lambda}
\end{equation} 

where $ \lambda=0.34 $ fm (full QCD) is the correlation length with the normalization fixed at $ T=\mu=0 $ by

\begin{equation}\label{D1E0}
D^{E}(0) +D_{1}^{E}(0)=\frac{\pi^{2}}{18} G_{2}
\end{equation} 
where $ G_{2} $ is the gluon condensate. The numerical value of $ G_{2} $ is determined by QCD sum rules with a large uncertainty as mentioned above~\cite{Shifman:1978by}:

\begin{equation}\label{G2}
G_{2}=0.012\pm0.006~{\text{GeV}}^{4}
\end{equation}
For the adapted parameter $ G_{2}=0.006 $ GeV$ ^{4} $, the  critical temperature turns out to be $ T=170 $ MeV, at zero chemical potential,~\cite{Simonov:2007jb}.
The generalization of the FCM at finite $ T $ and $ \mu $ provides expressions for the thermodynamic quantities where the leading contribution is given by the interaction of the single quark and gluon lines with the vacuum [called single line approximation (SLA)]. Within a few percent, in SLA the quark pressure for a single flavor is given as follows~\cite{Baldo:2008en,Simonov:2007jb,Komarov:2007zj}:
\begin{equation}\label{pq}
P_{q}/T^{4}=\frac{1}{\pi^{2}}[\varphi_{\nu}(\frac{\mu_{q}-V_{1}/2}{T})+\varphi_{\nu}(\frac{-\mu_{q}+V_{1}/2}{T})]~,
\end{equation} 
in which $ \nu=m_{q}/T $, and 
\begin{equation}\label{phinu}
\varphi_{\nu}(a)=\int_{0}^{\infty}du\frac{u^{4}}{\sqrt{u^{2}+\nu^{2}}} \frac{1}{(exp[\sqrt{u^{2}+\nu^{2}}-a]+1)}~,
\end{equation} 
and $ V_{1} $ is the large-distance static $ q\bar{q} $ potential:
\begin{equation}\label{v1}
V_{1}=\int_{0}^{1/T}d\tau(1-\tau T)\int_{0}^{\infty}d\chi\chi D_{1}^{E}(\sqrt{\chi^{2}+\tau^{2}})
\end{equation} 
The gluon contribution to the pressure is

\begin{equation}\label{pg}
P_{g}/T^{4}=\frac{8}{3\pi^{2}}\int_{0}^{\infty}d\chi\chi^{3}\frac{1}{exp(\chi+\frac{9V_{1}}{8T})-1}
\end{equation} 

Note that the potential $ V_{1} $ in Eq.~(\ref{v1}) does not depend on the chemical potential and this is partially supported by the lattice simulation at small chemical potential~\cite{Simonov:2007jb,Doring:2005ih}.

If confinement is dominated by nonperturbative contributions, the normalization $ D_{1}^{E}(0) $ in Eq.~(\ref{D1Ex}) can be identified with the term appearing in Eq.~(\ref{D1E0}) which has been denoted by the same symbol. Then, from Eqs.~(\ref{v1}), (\ref{D1Ex}) and (\ref{D1E0}) in the limit $ T\longrightarrow~0 $, we obtain

\begin{equation}\label{v1T0}
V_{1}(T=0)\leq\frac{\pi^{2}}{9}G_{2}\lambda^{3}
\end{equation}
However, other choices of $ V_{1} $ are possible, and these will be considered at the end of the result section.

The pressure in the quark-gluon phase can be written as~\cite{Simonov:2007jb,Komarov:2008vi}  

\begin{equation}\label{pqg}
P_{qg}=\sum_{i=u,d,s}P_{q}^{i}+P_{g}+\Delta\epsilon_{vac}
\end{equation}
where $ P_{q}^{i} $ and $ P_{g} $ are given in Eqs.~(\ref{pq}) and (\ref{pg}), respectively and
\begin{equation}\label{evac}
\Delta\epsilon_{vac}\approx-\frac{(11-2/3N_{f})}{32}\frac{G_{2}}{2}
\end{equation}

which corresponds to the difference of the vacuum energy density in the two phases, with $ N_{f} $ being the flavor number.

Other thermodynamic quantities in the quark-gluon phase can be derived in the standard way by using the relation 
\begin{equation} 
\epsilon=-P+\sum_{i=u,d,s}\mu_{i}n_{i}.
\end{equation}
As the weak decays ( $ d\leftrightarrow u+e+\bar{\nu}_{e}\leftrightarrow s $ ) should be taken into account in the quark-gluon matter, the electrons (neutrinos have enough time to leave the system) should be included, which are described by a noninteracting gas of massless fermions with
\begin{equation} 
P_{e}=\frac{\mu_{e}^{4}}{12\pi^{2}} ~~~~\rightarrow~~~~ \epsilon_{e}=\frac{\mu_{e}^{4}}{4\pi^{2}}
\end{equation}
Therefore, we will have
\begin{equation} 
P_{tot}=P+P_{e}~~~~~~~~~~~~\epsilon_{tot}=\epsilon+\epsilon_{e}~ 
\end{equation} 
 in the $ \beta $-stable quark-gluon matter. The relations between chemical potentials of the particles take the form
\begin{eqnarray}
\mu_{d}=\mu_{s}=\mu
\nonumber\\
\mu=\mu_{u}+\mu_{e}~
\end{eqnarray}

The charge neutrality condition implies ( $\frac{2}{3}n_{u}-\frac{1}{3}n_{d}-\frac{1}{3}n_{s}-n_{e}=0  $ ) and so the system can be characterized by one independent variable, that is, the baryon number density $ \rho_{B}=\frac{1}{3}(n_{u}+n_{d}+n_{s}) $.

\section{Results}

\subsection{Hadron-quark hybrid EOS}\label{IV-A}
We study the hadron-quark phase transition in order to obtain the EOS of a hybrid star. We consider the Maxwell construction by assuming a first-order hadron-quark phase transition. Maxwell construction is a sharp phase transition from neutral hadronic matter to homogeneous neutral quark matter. Both hadron and quark phases are in $ \beta $ equilibrium and also satisfy charge neutrality, separately. Each phase is considered to be a one-component system controlled by the baryonic density or equivalently a baryonic chemical potential because of the requirement of the charge neutrality in Maxwell construction. By imposing the conditions of thermal, mechanical, and one-component chemical equilibrium at zero temperature, the transition point in the Maxwell construction is identified as
\begin{equation}\label{p} 
P_{1}(\mu_{B})=P_{2}(\mu_{B})
\end{equation}
where the indices 1 and 2 stand for the hadronic and quark phases, respectively. Equation~(\ref{p}) implies that Maxwell construction corresponds to constant pressure in the density interval between two phases.  $\mu_{B}  $ stands for the baryon chemical potential in each phase ($\mu_{B1}=\mu_{p} +\mu_{n}  $ and $ \mu_{B2}=\mu_{u} +\mu_{d} +\mu_{s} $). At the interface between the two phases, the baryon chemical potential $ \mu_{B} $ is continuous while the electron chemical potential $ \mu_{e} $ jumps in Maxwell construction. One can consider Maxwell construction as a limiting scenario where the surface tension is large.

\begin{figure*}[htb]
	\vspace{-0.50cm}
	\resizebox{0.45\textwidth}{!}{\includegraphics{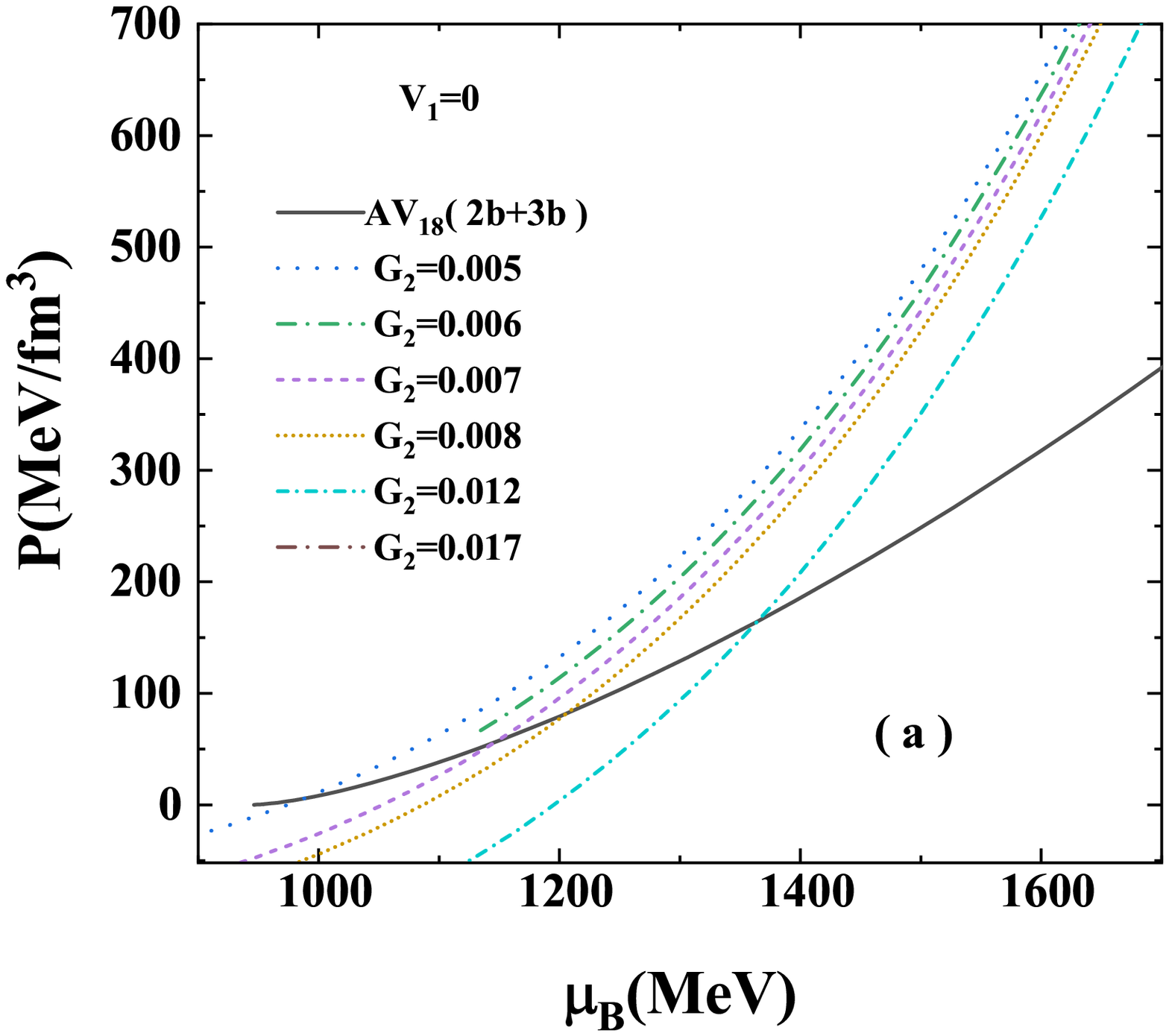}}
	\resizebox{0.45\textwidth}{!}{\includegraphics{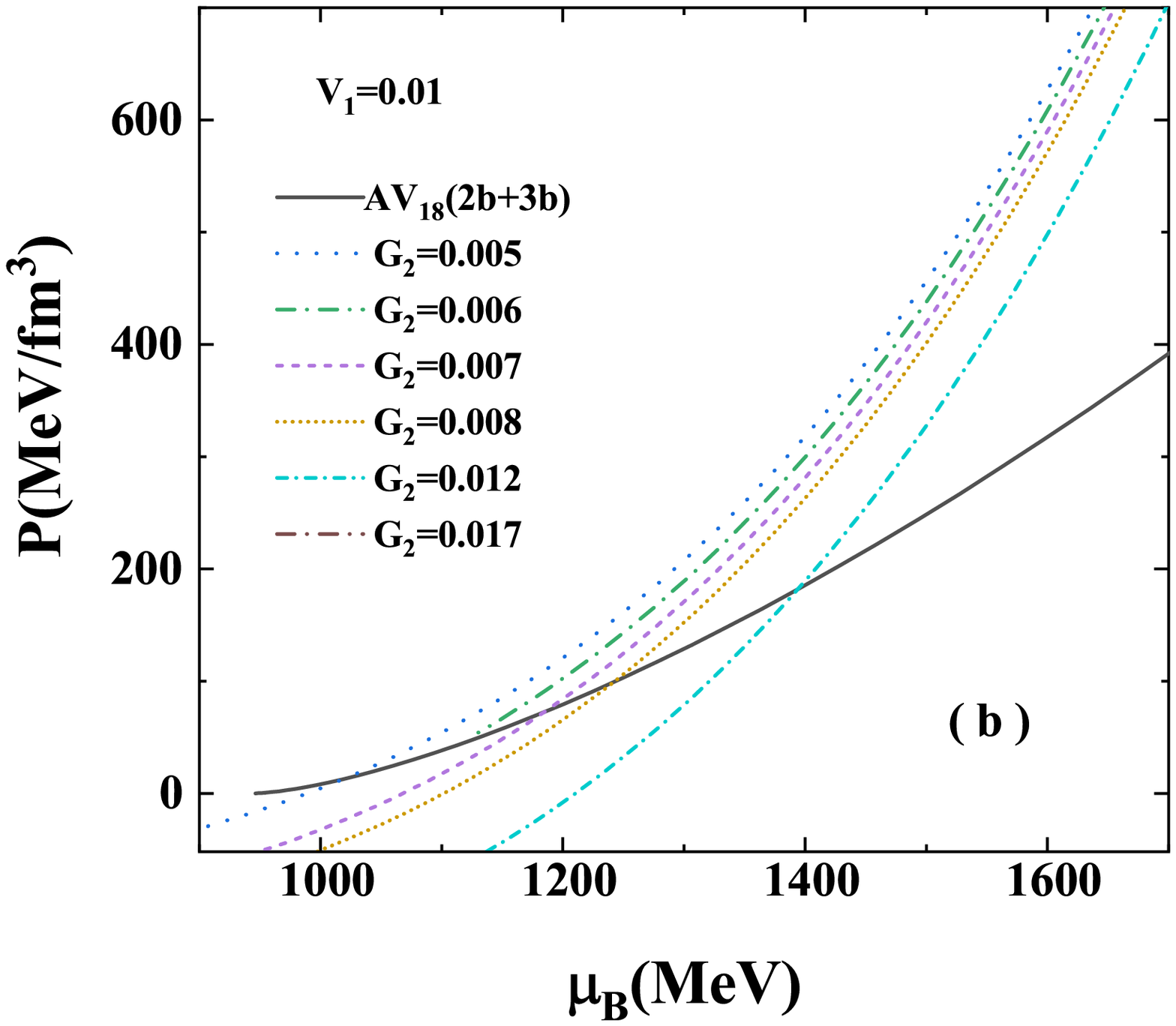}}
		\resizebox{0.45\textwidth}{!}{\includegraphics{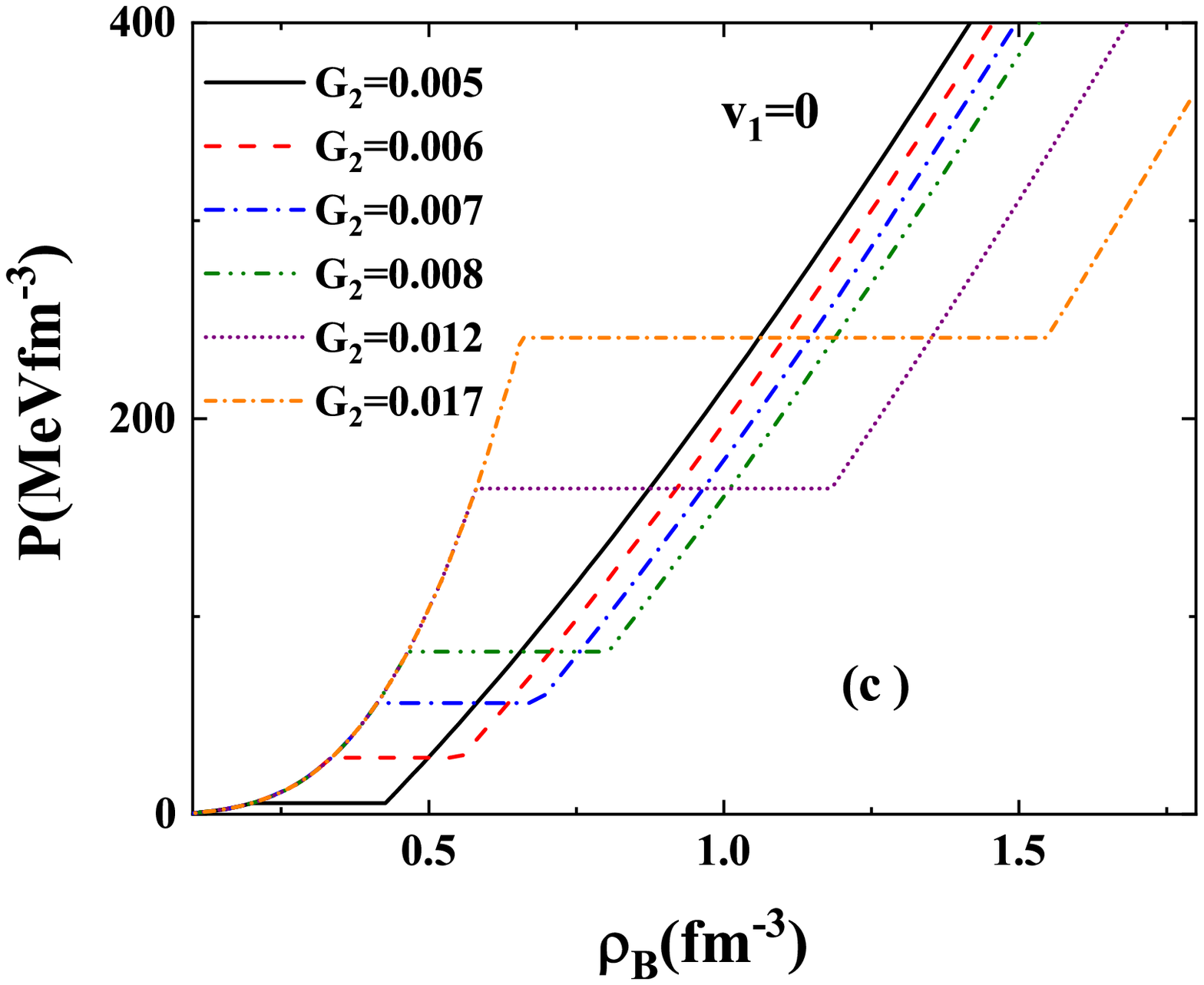}}
	\resizebox{0.44\textwidth}{!}{\includegraphics{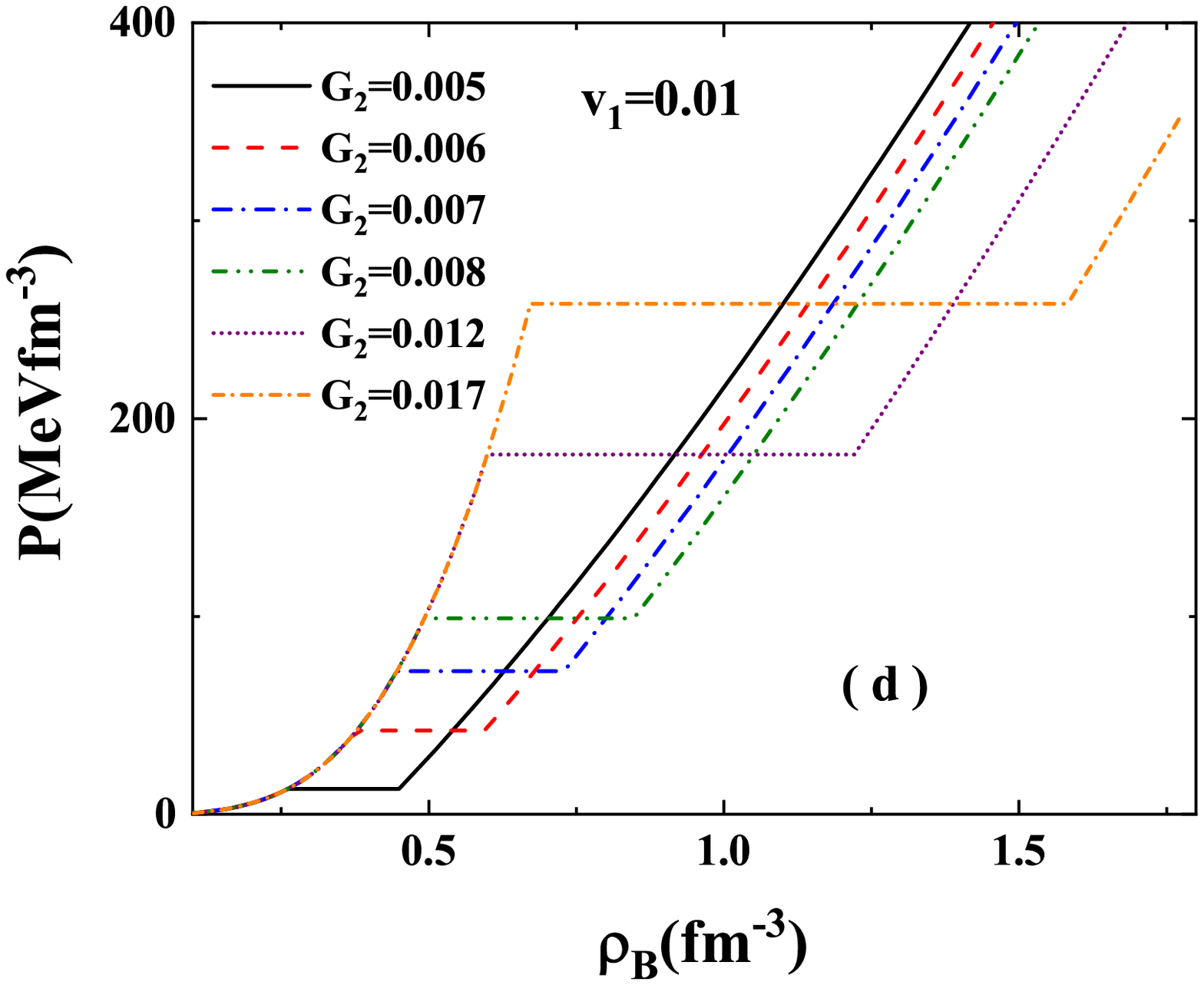}}
	\begin{center}
		\caption{{\small (a),(b) : Pressure vs baryon number density for $ \beta $-stable nuclear matter within the LOCV method supplemented with the TBF and $ \beta $- stable quark matter in FCM model with (a) $ V_{1}=0 $ (b) $ V_{1} =0.01$ GeV and several values for the gluon condensate $ G_{2} $ (in GeV$ ^{4} $)}. (c),(d) : The hadron-quark hybrid EoSs in Maxwell construction with $ q\bar{q} $ potential (c) $ V_{1}=0 $ (d) $ V_{1}=0.01 $ GeV and several choices of gluon condensate $ G_{2} $ (in GeV$ ^{4} $) combined with the LOCV supplemented by TBF.
		} \label{fig1} 
	\end{center}
\end{figure*}

 In Fig.~\ref{fig1}, the pressure $ P $ as a function of the baryon chemical potential $ \mu _{B} $ for baryonic and quark matter phases in $ \beta $ equilibrium is shown, and also the hybrid EOSs (Pressure $ P $ vs baryon density $ \rho_{B} $) are displayed. In Fig.~\ref{fig1}(a) [\ref{fig1}(b)], we show the results obtained using $ V1=0 $ [$ V1=0.01 $ GeV] $ q\bar{q} $ potential [according to the constant obtained in Eq.~(\ref{v1T0})]. In both panels, the solid black line represents the EOS of the nuclear matter with $ AV_{18} $ potential supplemented by TBF in LOCV formalism, and other lines represent the EOS of quark-gluon phase within the FCM with several choices of parameter sets. It is worth noting that the chosen values of $ G_{2} $ give the values of the critical temperature in a range between $ T\approx150 $ to $ 200 $ MeV. The transition point in the $ V_{1}=0.01 $ GeV case, is shifts slightly to higher values of chemical potential, and also the baryon density, compared to the case in which $ V_{1}=0 $. We notice that the crossing point is significantly affected by the choice of the gluon condensate, $ G_{2} $. With an increase the value of the gluon condensate $ G_{2} $, the transition point shifted to higher values of chemical potential. However, the exact value depends also on the stiffness of the baryonic EOS at those densities. The onset of phase transition is around $ 2\rho_{0} $ ($  \rho_{0}=0.16 $ fm$ ^{-3} $).
  In Fig.~\ref{fig1}(c) [\ref{fig1}(d)], the hybrid EOS in Maxwell construction is displayed for several cases discussed. The result obtained with $ V1=0 $ [$ V1=0.01 $ GeV] is displayed in Fig.~\ref{fig1}(c) [\ref{fig1}(d)]. Below the plateau the $ \beta $-stable hadronic EOS governs the star while in densities higher than the ones characterized by the plateau, the stellar matter is in the $ \beta $-stable quark matter phase. 
  
  It is clear that the width of the plateau is related to the values of $ q\bar{q} $ potential, $ V_{1} $, gluon condensate $ G_{2} $ and the baryonic EOS. With increasing values of $ G_{2} $, and $ V_{1} $, the width is extended. As the width of the plateau increases, the discontinuity in energy density between the two phases increases, which in turn causes the instability of the HS. (We will refer to this point later.)

\subsection{Hybrid star structure }\label{IV-B}

The structure of a hybrid star is calculated by numerical integration of the well-known hydrostatic equilibrium equations of Tolman-Oppenheimer-Volkoff (TOV) . The  EOS of the star is the fundamental input of the  TOV equations:
\begin{eqnarray}
\frac{dP(r)}{dr} =&&-\frac{GM(r)\epsilon (r)}{c^{2} r^{2} } (1+\frac{P(r)}{\epsilon (r)} )(1+\frac{4\pi r^{3} P(r)}{M(r)c^{2} } )
\nonumber\\
&&\times(1-\frac{2GM(r)}{rc^{2} } )^{-1},
\end{eqnarray}
\begin{equation}
\frac{dM(r)}{dr} =\frac{4\pi \epsilon (r)r^{2} }{c^{2} },
\end{equation}
in which $\epsilon (r)$ is the total energy density, $ M(r) $ is the star mass within radius $ r $, $ c $ is the speed of light, and $G$ denotes the gravitational constant. 

The hybrid EOS in Maxwell construction, with constant pressure in the transition region, is taken from the calculations discussed above. For the description of the NS crust, we use the Harrison-Wheeler (HW) EOS. The effects of different crustal EOSs on the structure of hybrid star  are studied in Sec.~\ref{IV-C}.

\begin{figure*}[htb]
	\vspace{-0.50cm}
	\resizebox{0.45\textwidth}{!}{\includegraphics{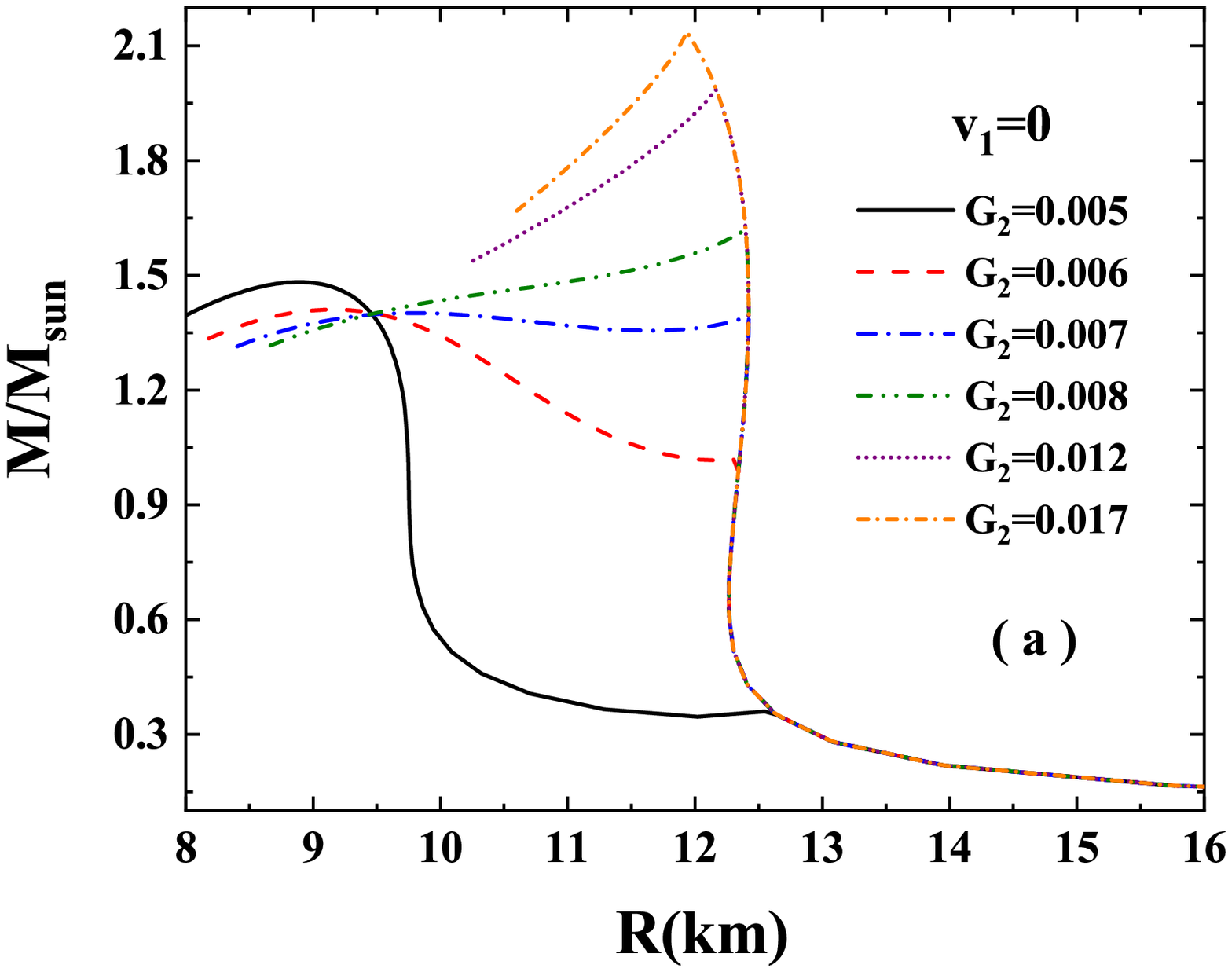}}
	\resizebox{0.44\textwidth}{!}{\includegraphics{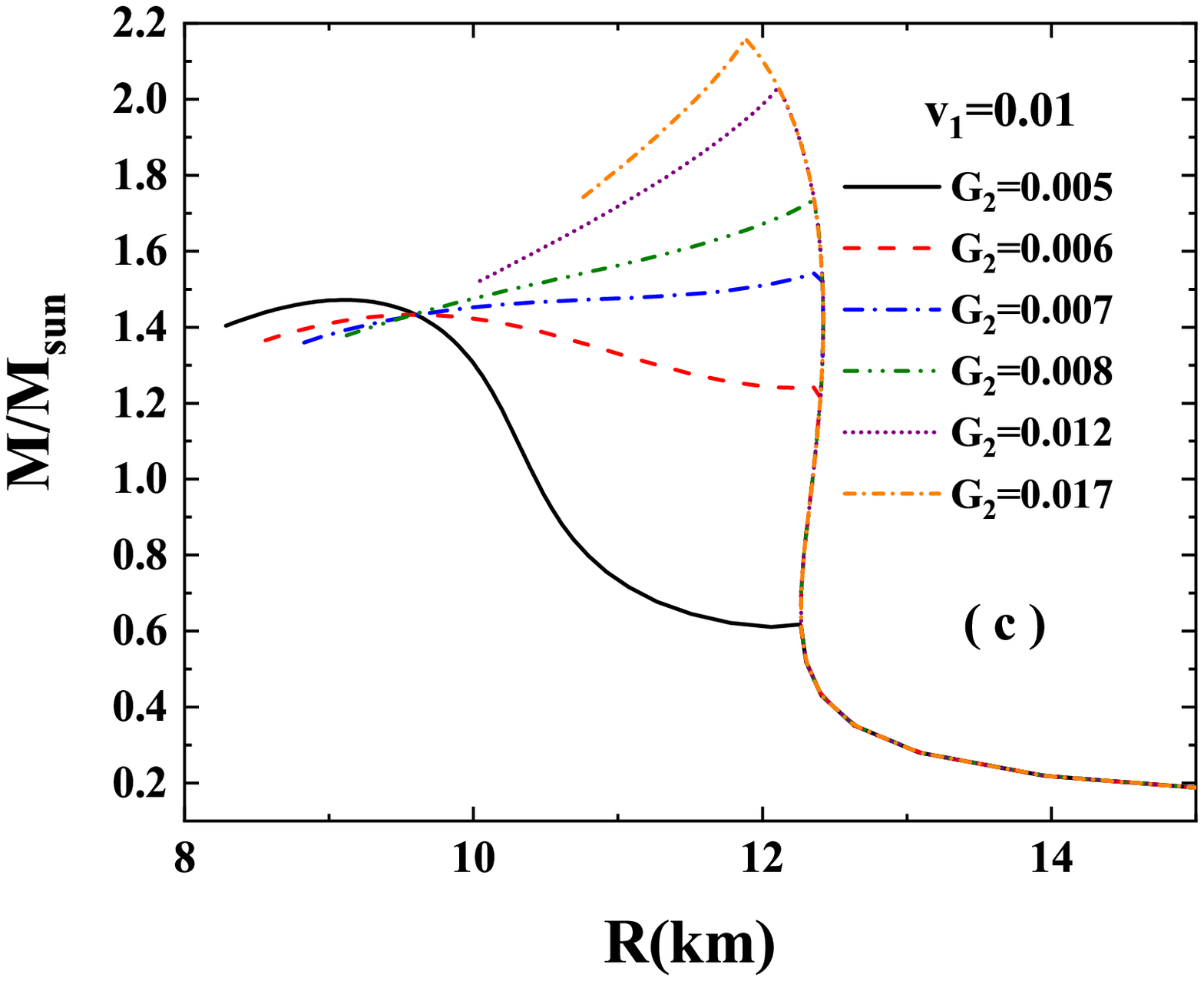}}
	\resizebox{0.45\textwidth}{!}{\includegraphics{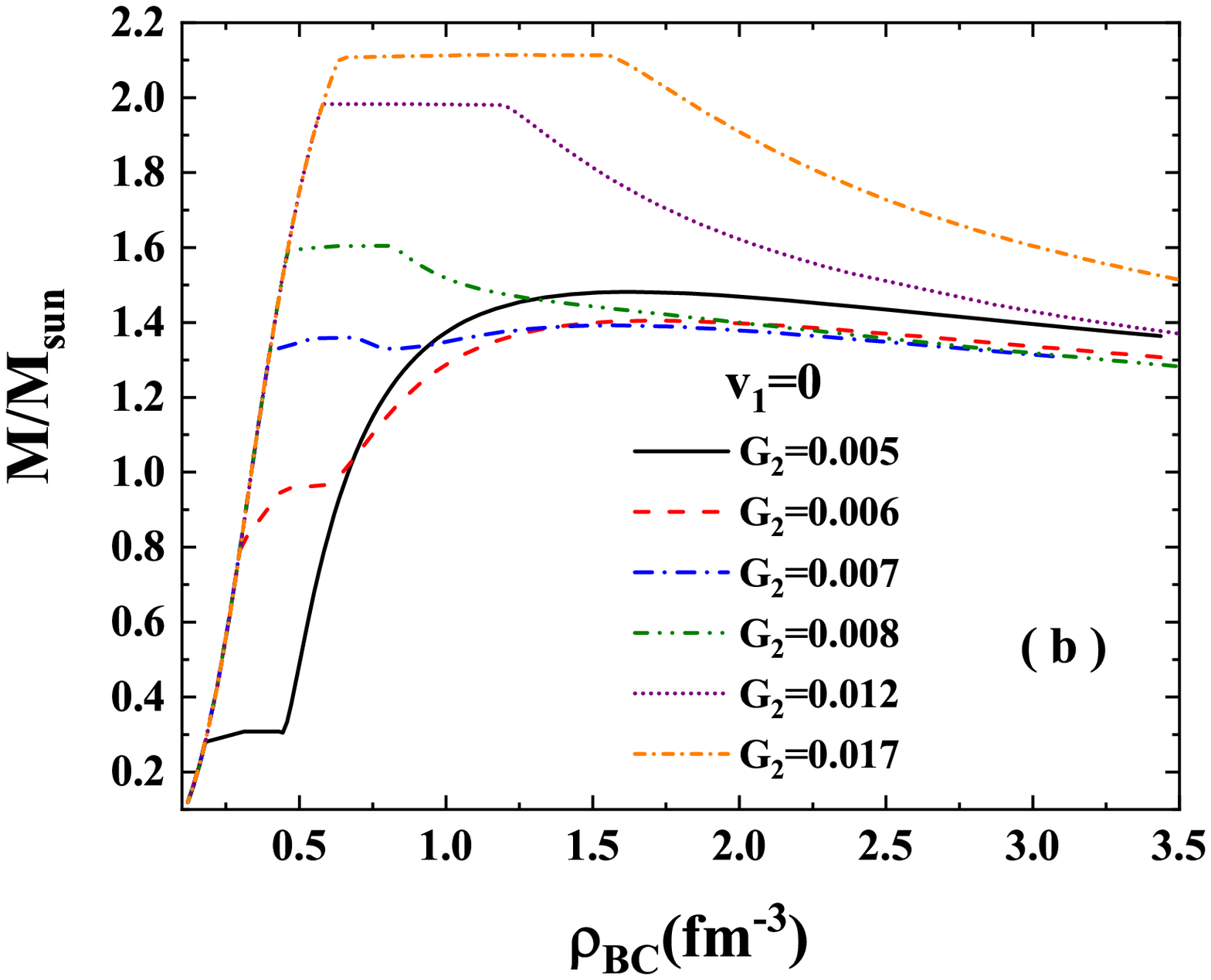}}
	\resizebox{0.44\textwidth}{!}{\includegraphics{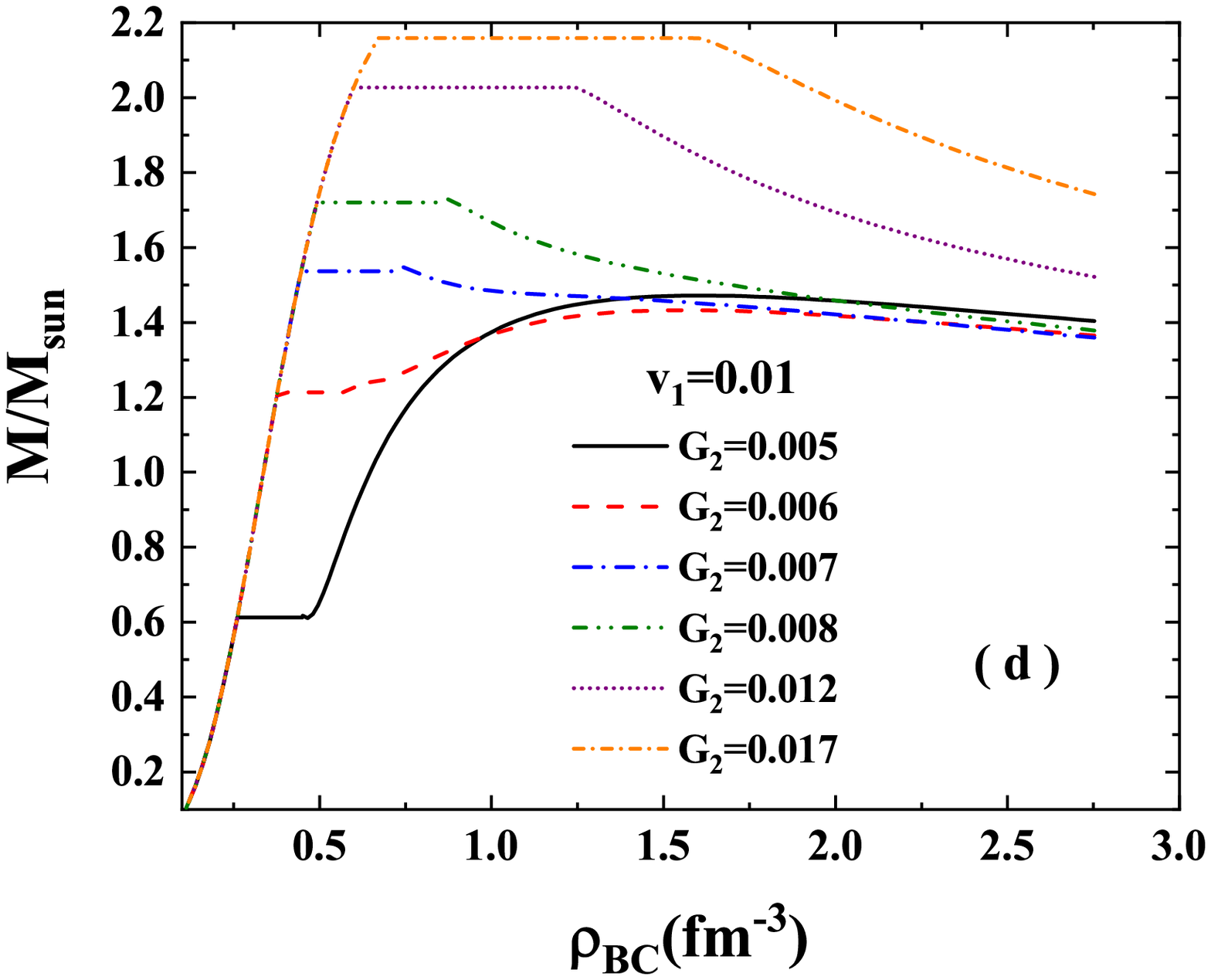}}
	\begin{center}
		\caption{{\small  (a),(b)  : The gravitational HS masses vs (a) radius (b) central baryon density of the star  with the $ q\bar{q} $ potential $ V_{1}=0 $ and several choices of gluon condensate $ G_{2} $ (GeV$ ^{4} $) combined with $ AV_{18} $ supplemented by TBF. (c),(d) : Same as (a),(b) but with the the $ q\bar{q} $ potential $ V_{1}=0.01  $ GeV. } \label{fig2}}
	\end{center}
\end{figure*}

In Fig.~\ref{fig2}(a) [\ref{fig2}(b)], we display the mass-radius [mass-central density] for hybrid stars with $ q\bar{q} $ potential $ V_{1}=0 $ for several choices of gluon condensate, $ G_{2} $. Figure~\ref{fig2}(c) [\ref{fig2}(d)] is the same as the previous case, but for $ q\bar{q} $ potential $ V_{1}=0.01 $ GeV. 
By looking at Fig.~\ref{fig2}(a) [\ref{fig2}(b)], we find that the maximum mass of HS spans over a range between $ 1.4M_{\odot} $ and $ 2.16M_{\odot} $ depending on the values of the gluon condensate $ G_{2} $ and $ q\bar{q} $ potential $ V_{1} $. The HS with the maximum mass of $ 2.13M_{\odot} $ is predicted for $V_{1}=0 $ and the gluon condensate, $ G_{2}=0.017 $ GeV$ ^{4} $. By switching on the value of $ q\bar{q} $ potential, $ V_{1} $, as displayed in Fig.~\ref{fig2}(c) [\ref{fig2}(d)], we observe a trend similar to the case of $ V_{1}=0 $. The value of maximum mass slightly increases in the case $ V_{1}=0.01 $ GeV with respect to the case $ V_{1}=0 $. In the case $ V_{1}=0.01 $ GeV, the maximum mass of $ 2.03M_{\odot} $ ($ 2.16M_{\odot} $) is calculated for $ G_{2}=0.12 $ GeV$ ^{4} $ ($ G_{2}=0.17 $ GeV$ ^{4} $). However, in the mentioned cases, with the maximum masses compatible with observations, the HSs are unstable. The instability manifests itself as a cusp in the mass-radius curve , which in turn is due to the large discontinuity in the energy density in the phase transition region. A stable HS with a pure quark core is predicted only for small values of $ G_{2} $ around less than $ G_{2}=0.07 $ GeV with the maximum mass of about $ 1.4M_{\odot} $, which are hardly in agreement with the observations. It is worth noting that an "acceptable" EOS must give a maximum mass around $ 2M_{\odot} $. By increasing the value of the gluon condensate, $ G_{2} $, the value of the maximum mass increases, up to about $ 2.16M_{\odot} $; however, the stability of a pure quark core is lost. The results are summarized in Table~\ref{t1}. As seen in Table~\ref{t1}, in the cases with energy density discontinuity around 300 MeV fm$^{-3} $, the HS with a pure quark core is stable. For higher values of  discontinuity in the energy density, the HS becomes unstable. Therefore, generally speaking, the FCM model with very low values of $ q\bar{q} $ potential,$ V_{1} $, as predicted by Eq.(~\ref{v1T0}),  gives a maximum value of mass higher than $ 2M_{\odot} $ for large values of the gluon condensate (around $ G_{2}=0.012 $ GeV$ ^{4} $), and the star becomes unstable as soon as the onset of the quark phase. A stable pure quark core is predicted in low values of the gluon condensate with a maximum mass around $ 1.4M_{\odot} $.

\begin{figure*}[htb]
	\vspace{-0.50cm}
	\resizebox{0.45\textwidth}{!}{\includegraphics{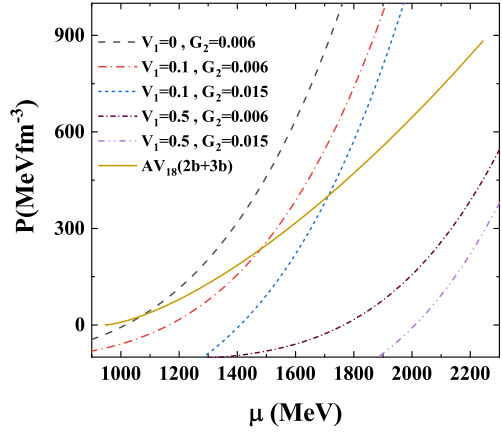}}
	
	\begin{center}
		\caption{{\small Pressure vs baryon chemical potential, for different values of gluon condensate, $ G_{2} $ (GeV$ ^{4} $), and $ q\bar{q} $ potential, $ V_{1} $ (GeV). } \label{fig3}} 
	\end{center}
\end{figure*}

As we mentioned before, lattice calculation determines the value of gluon condensate to be $ G_{2}=0.006 $ GeV$ ^{4} $ at critical temperature and $ \mu=0 $, while up to now, our calculations predict the maximum value of the HS mass to be around $ 1.4 M_{\odot} $, which is far from the observational data. This puts a serious constraint on the value of the gluon condensate. However, this prediction is obtained for the very low value of long-distance static $ q\bar{q} $ potential $ V_{1} $ arising from Eq.(~\ref{v1T0}). Other choices are possible. If Eq.~(\ref{D1Ex}) is assumed to be valid only at long range, while Eq.~(\ref{D1E0}) is a true short-range relationship, then in this case the parameter $ D_{1}^{E}(0) $ in the two equations cannot be identified and may correspond to two different numerical values, and therefore the value of $ V_{1} $ must be considered as an independent parameter~\cite{Baldo:2008en}. In the comparison with lattice calculations~\cite{Simonov:2007xc}, one finds a value of $ V_{1}=0.5 $ GeV at the critical temperature and for $ \mu=0 $. Besides that, the assumption of the independence of $ V_{1} $ on $ \mu $ can be questioned, and in any case, the value of this parameter is quite uncertain at high densities and low temperature ~\cite{Baldo:2008en}. We have therefore varied the strength of $ V_{1} $ from small values considered previously up to 0.5 GeV. The results for the EOS are reported in Fig~\ref{fig3} for different values of $ q\bar{q} $ potential $ V_{1} $. One can see that the hadron-quark phase transition is shifted to higher values of the chemical potential and hence of the densities. Actually, for $ q\bar{q} $ potential $ V_{1}=0.1 $ GeV, the phase transition occurs, while for $ V_{1}=0.5 $ GeV, there are no crosses between hadronic and quark matter EOSs,  and therefore the quark phase is irrelevant for NS physics.

In order to obtain the probable stable HS with higher maximum masses, we have carried out the calculation for larger values of $ q\bar{q} $ potential; $ V_{1}=0.05, 0.07,0.09,0.1,0.12 $ GeV.

\begin{figure*}[htb]
	\vspace{-0.50cm}
	
	\resizebox{0.325\textwidth}{!}{\includegraphics{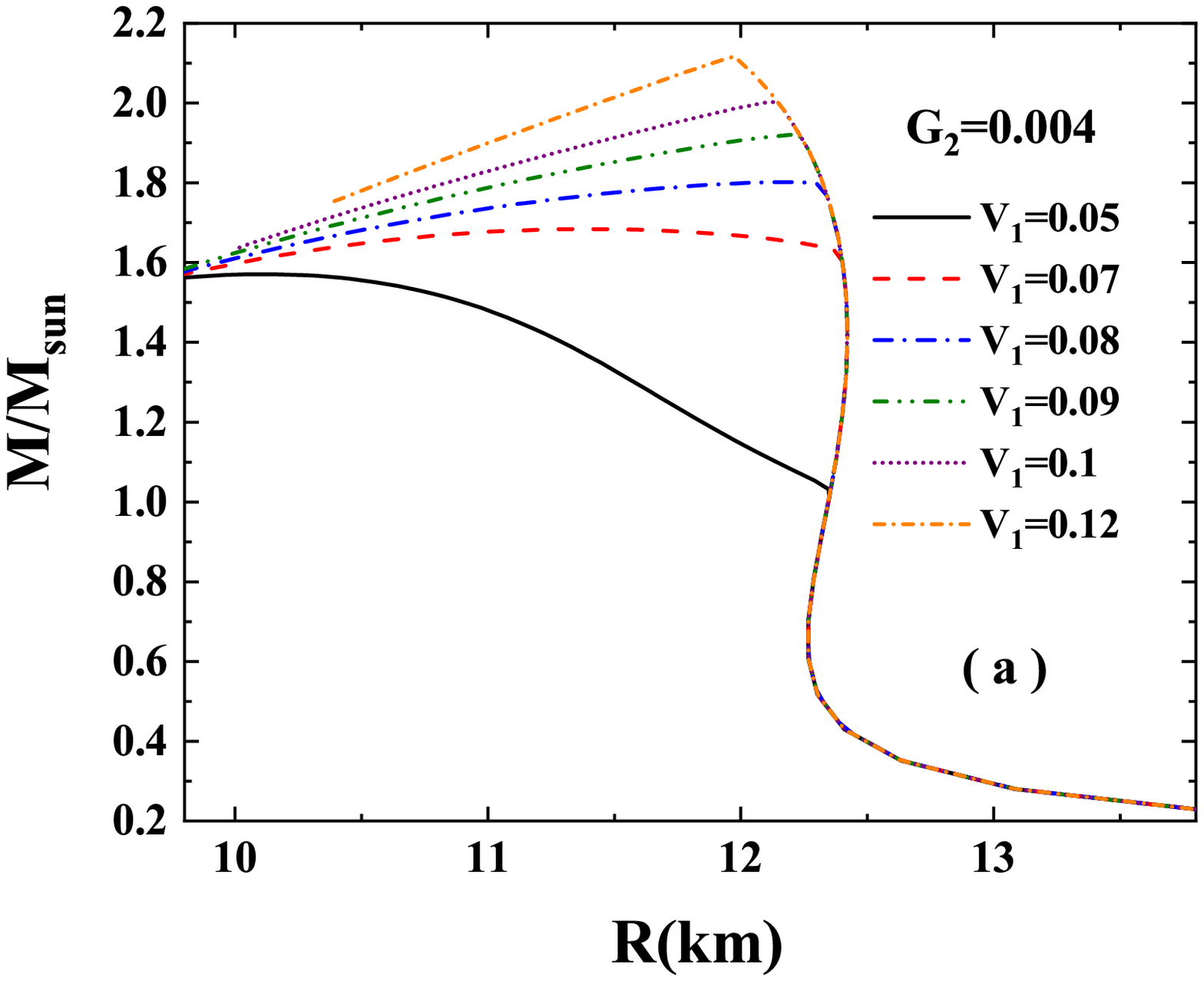}}
	\resizebox{0.325\textwidth}{!}{\includegraphics{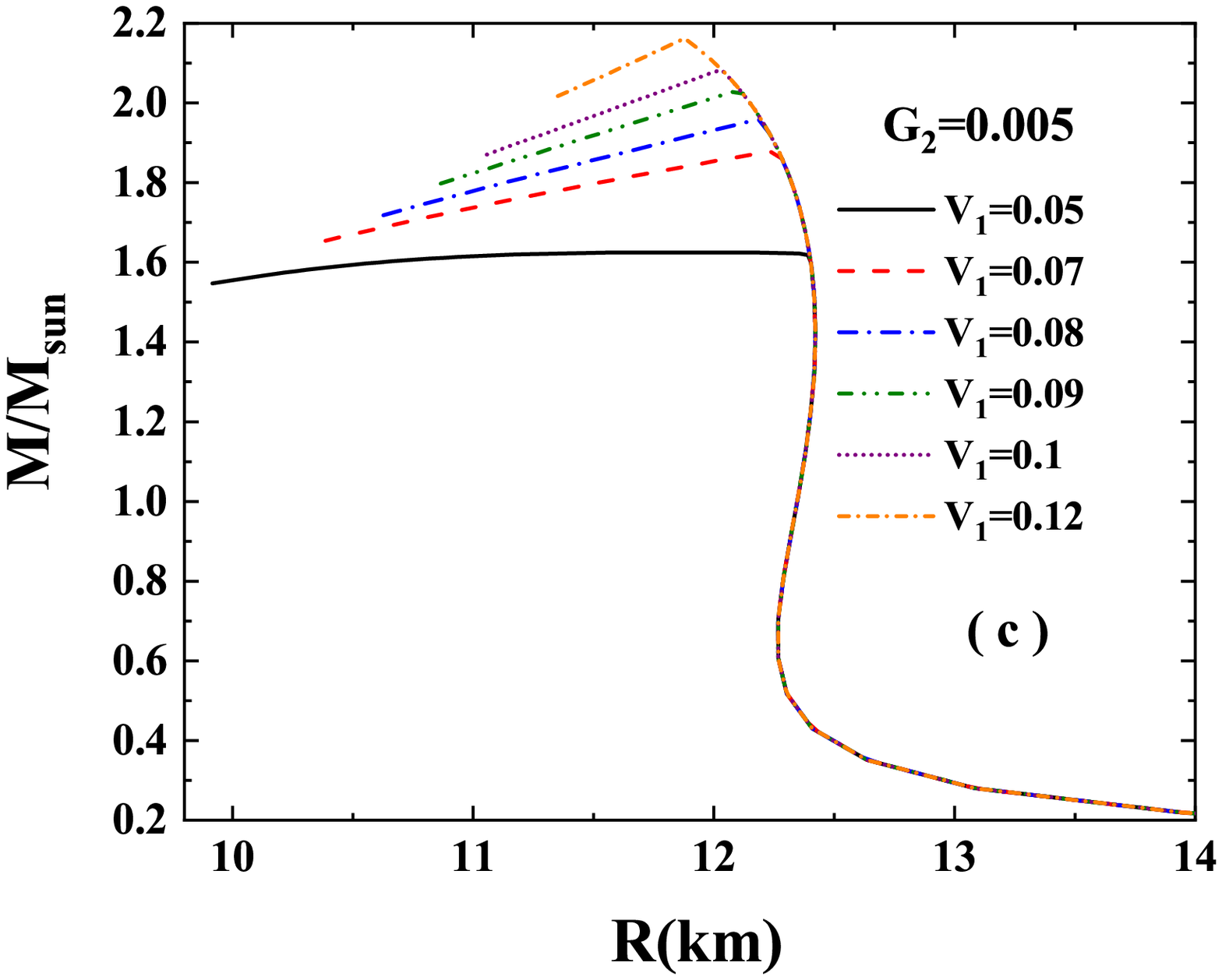}}
	\resizebox{0.325\textwidth}{!}{\includegraphics{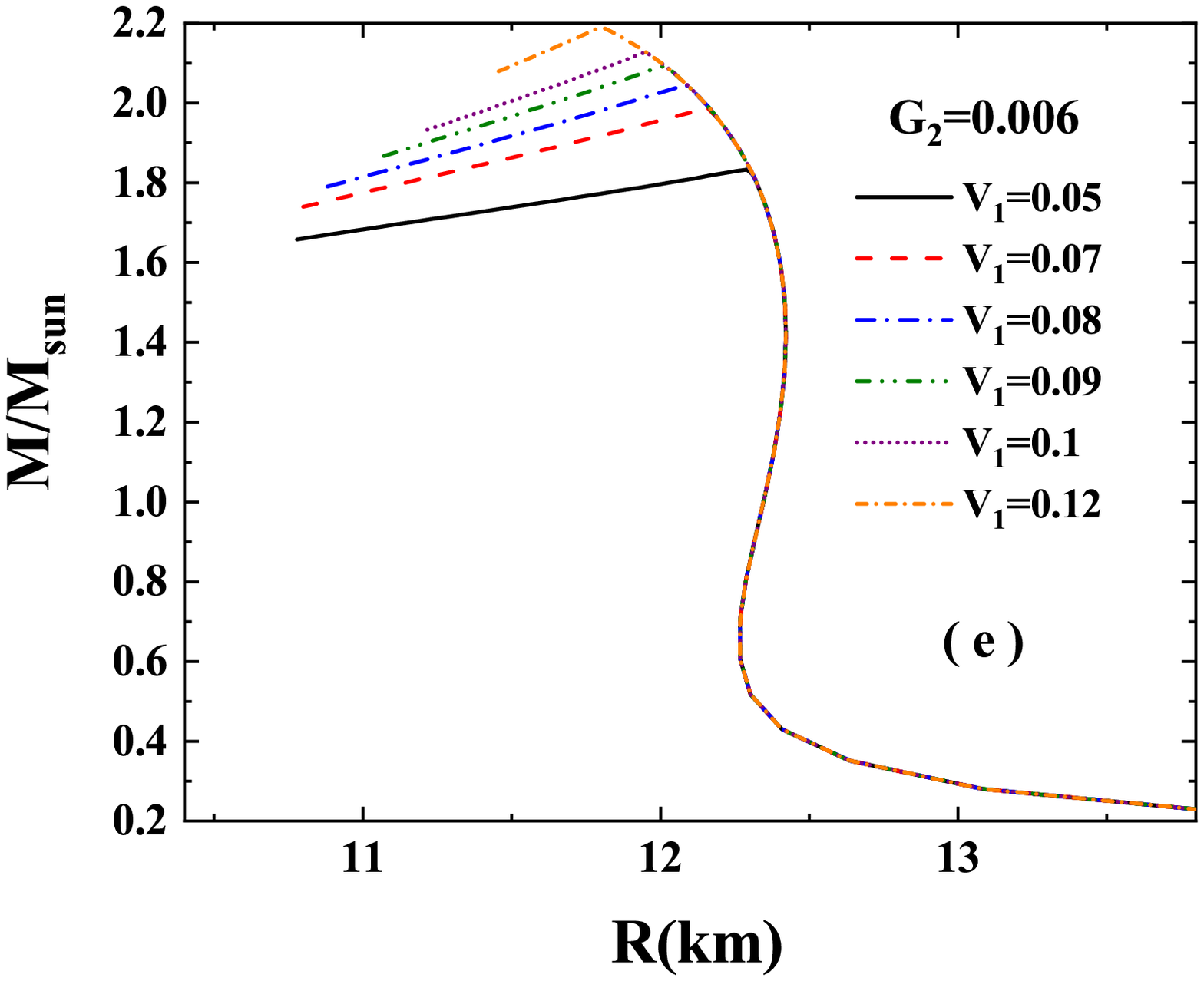}}
	\resizebox{0.325\textwidth}{!}{\includegraphics{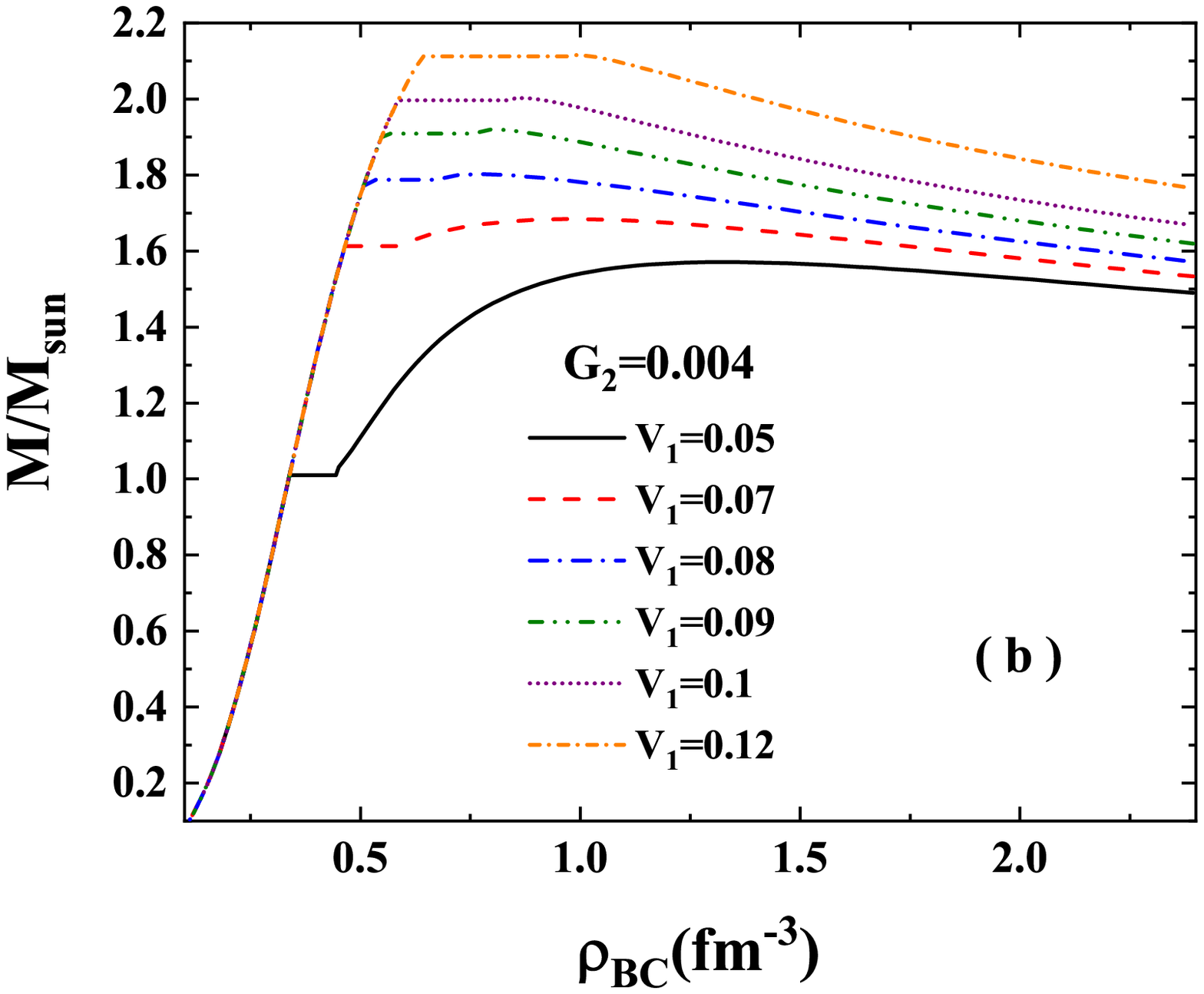}}
	\resizebox{0.325\textwidth}{!}{\includegraphics{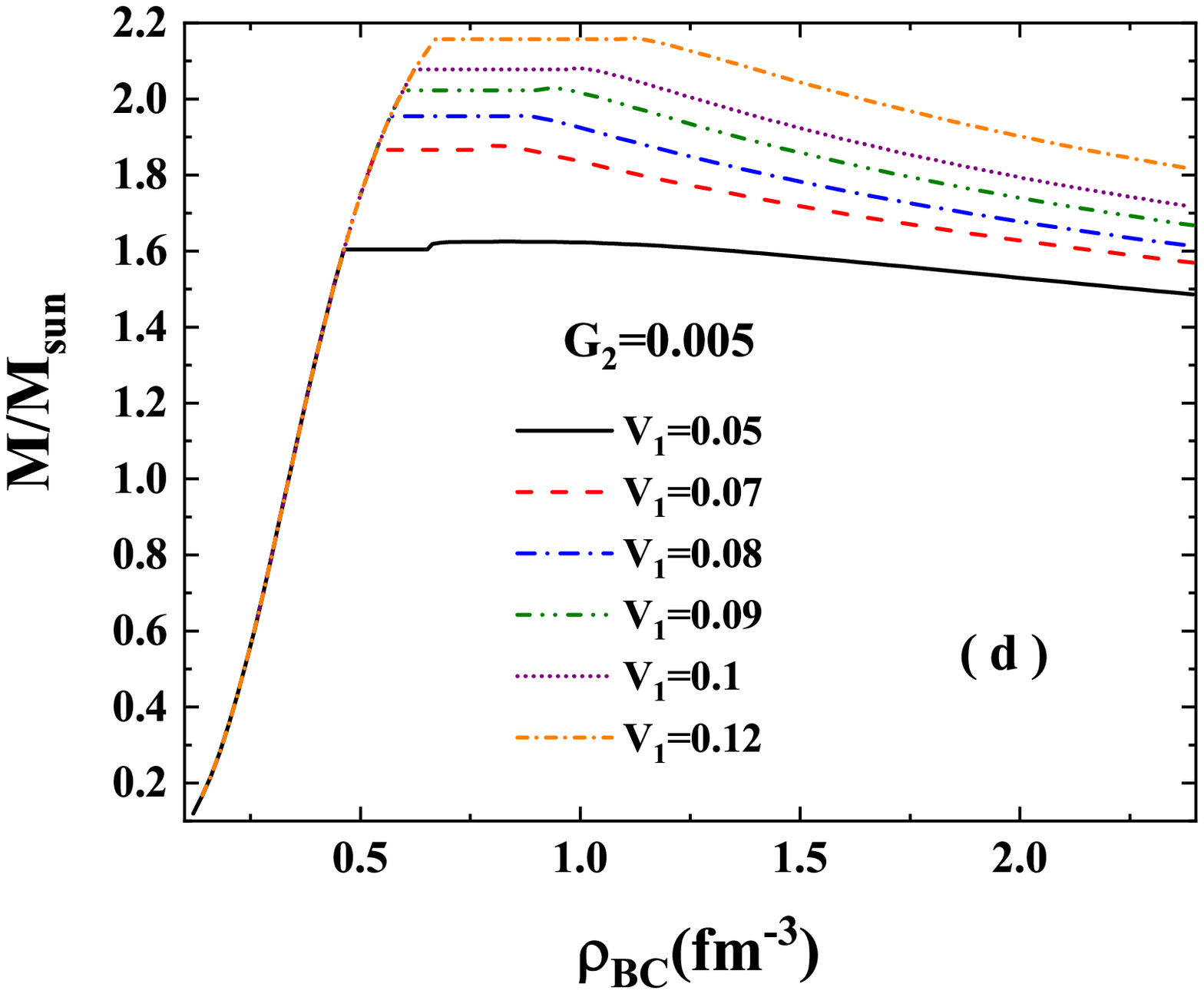}}
	\resizebox{0.325\textwidth}{!}{\includegraphics{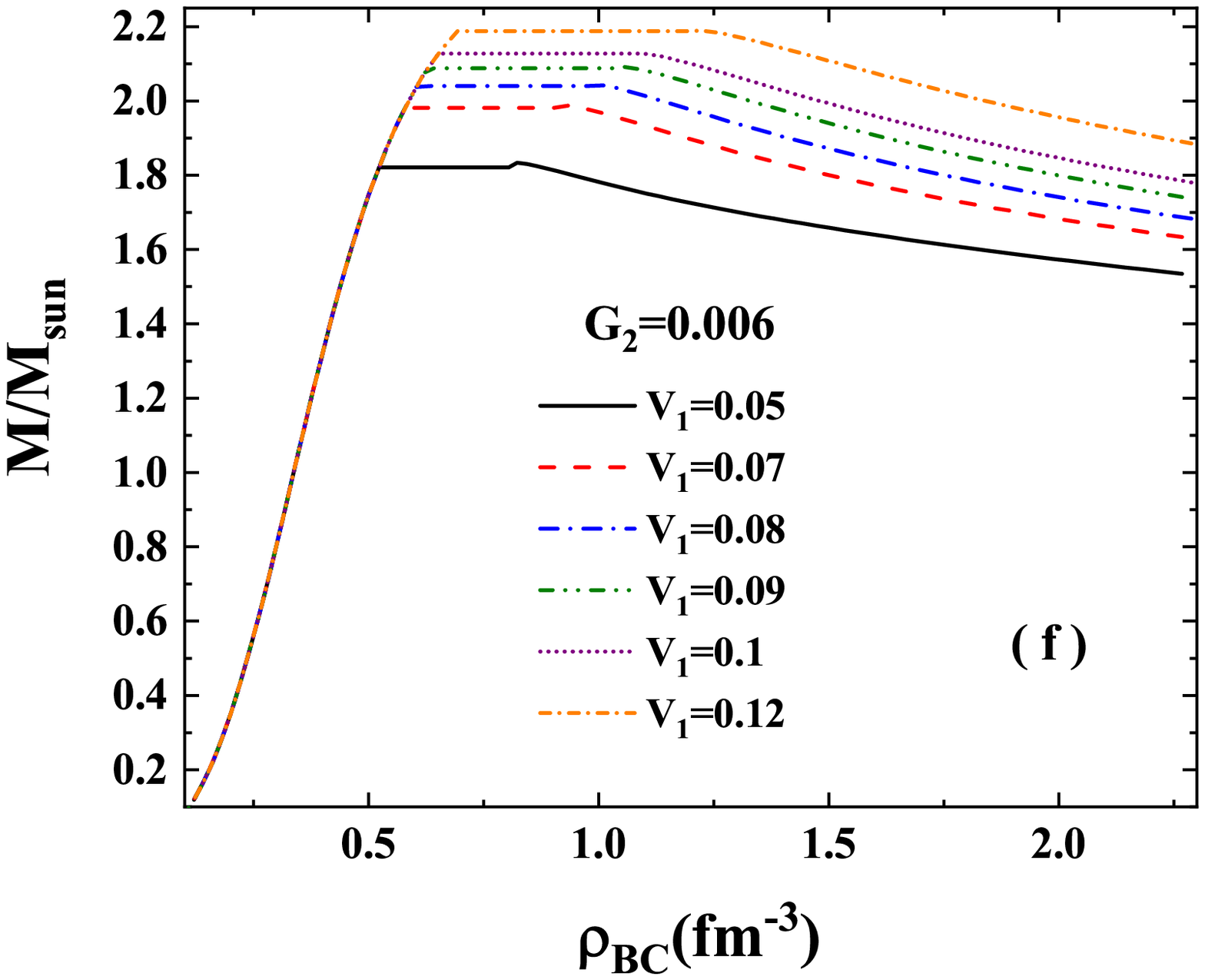}}
	\begin{center}
		\caption{{\small (a),(b) : The gravitational HS masses vs (a) radius (b) central baryon density of the star  with the gluon condensate  $ G_{2}=0.004 $ GeV$ ^{4} $ and several $ q\bar{q} $ potential $V_{1}  $ (in GeV). (c),(d) : Same as (a),(b),  but with the gluon condensate  $ G_{2}=0.005 $ GeV$ ^{4} $.   Panel (e),(f) : Same as (a),(b)  but with the gluon condensate  $ G_{2}=0.006 $ GeV$ ^{4} $ }\label{fig4} } 
	\end{center}
\end{figure*}

In Fig.~\ref{fig4}, we display the effect of increasing the $ q\bar{q} $ potential $ V1 $ on the maximum mass of the HS. In Fig.~\ref{fig4} (a) [\ref{fig4} (b)], the mass-radius [mass-central density] of the HS with a gluon condensate value of $ G_{2}=0.004 $GeV$ ^{4} $ and several values of $ q\bar{q} $ potential $ V_{1} $ are displayed. Figures~\ref{fig4} (c) [\ref{fig4} (d)] and \ref{fig4} (e) [\ref{fig4} (f)] are the same as the previous case, but with the gluon condensate values of $ G_{2}=0.005 $ GeV$ ^{4} $ and $ G_{2}=0.006 $ GeV$ ^{4} $. By increasing the $ q\bar{q} $ potential, $ V_{1} $, the maximum mass increases, and simultaneously, the HS becomes unstable. For the case $ G_{2}=0.004 $ GeV$ ^{4} $, as seen in Fig.~\ref{fig4} (a) [\ref{fig4} (b)], a stable HS is predicted up to $ q\bar{q} $ potential $ V_{1}=0.09 $ GeV with the maximum mass $ 1.92M_{\odot} $. For larger $ V_{1} $, the HS become unstable. As seen in Fig.~\ref{fig4} (c) [Fig.~\ref{fig4} (d)], for the case $ G_{2}=0.005 $ GeV$ ^{4} $, the stable HS predicted up to $ q\bar{q} $ potential, $ V_{1}=0.09 $ GeV with the maximum mass $ 2.03M_{\odot} $. For larger $ V_{1} $, the pure quark core becomes unstable. As seen in Fig.~\ref{fig4} (e) [\ref{fig4} (f)] for the case $ G_{2}=0.006 $ GeV$ ^{4} $, a stable HS is predicted up to $ q\bar{q} $ potential, $ V_{1}=0.08 $ GeV with the maximum mass $ 2.04M_{\odot} $. For larger $ V_{1} $, the star becomes unstable. The results are summarized in Table~\ref{t2}. The  discontinuity in the energy density in stable HSs is around 500 MeV fm$ ^{-3} $. By increasing the value of $q\bar{q} $ potential, $ V_{1} $, to higher than 0.09 GeV, the value of the maximum mass is shifted higher than $ 2M_{\odot} $ and simultaneously the stability of the star is lost for all values of gluon condensates $ G_{2} $. The results are summarized in Table~\ref{t3}. 

\begin{figure*}[htb]
	\vspace{-0.50cm}
	\resizebox{0.45\textwidth}{!}{\includegraphics{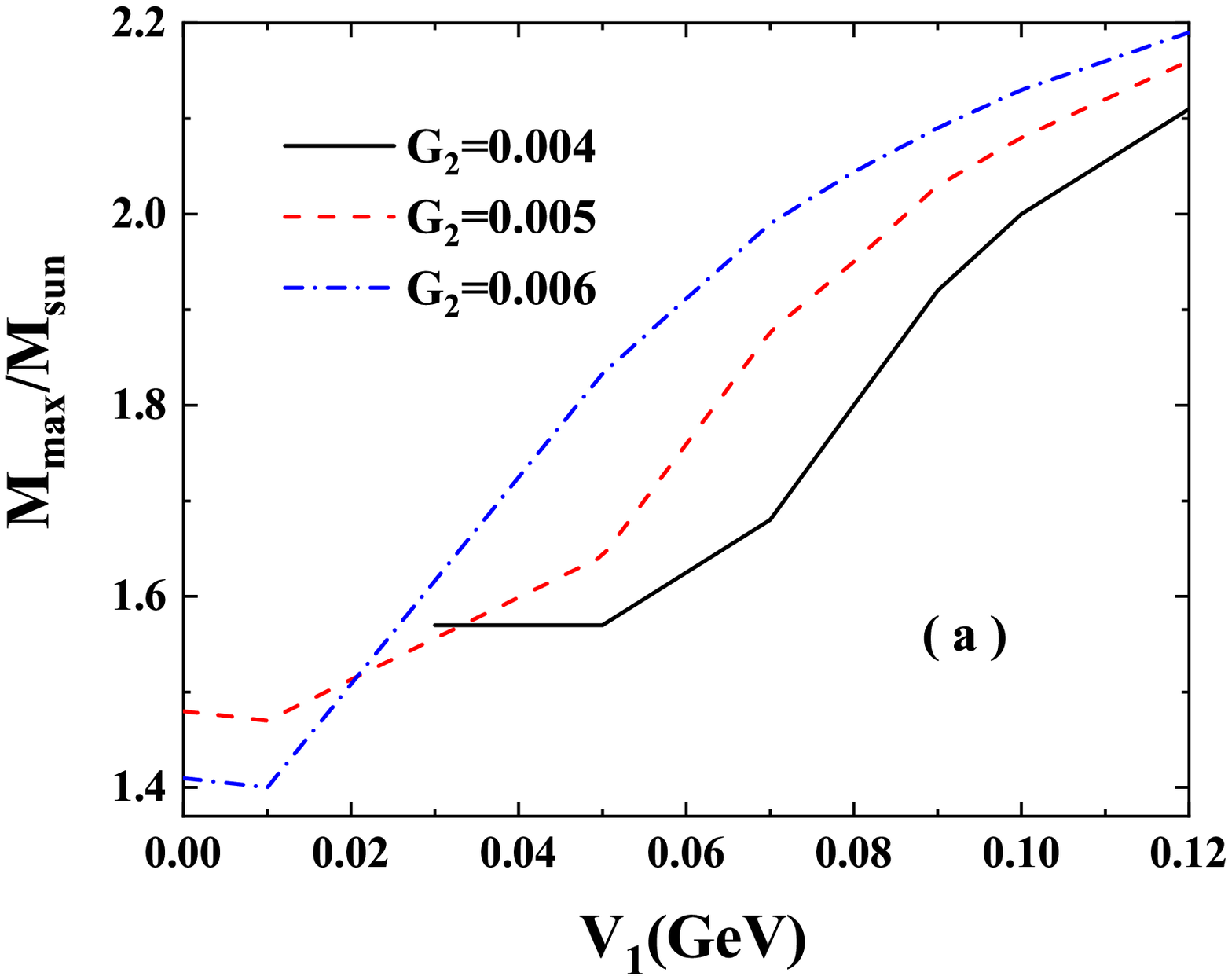}}
	\resizebox{0.45\textwidth}{!}{\includegraphics{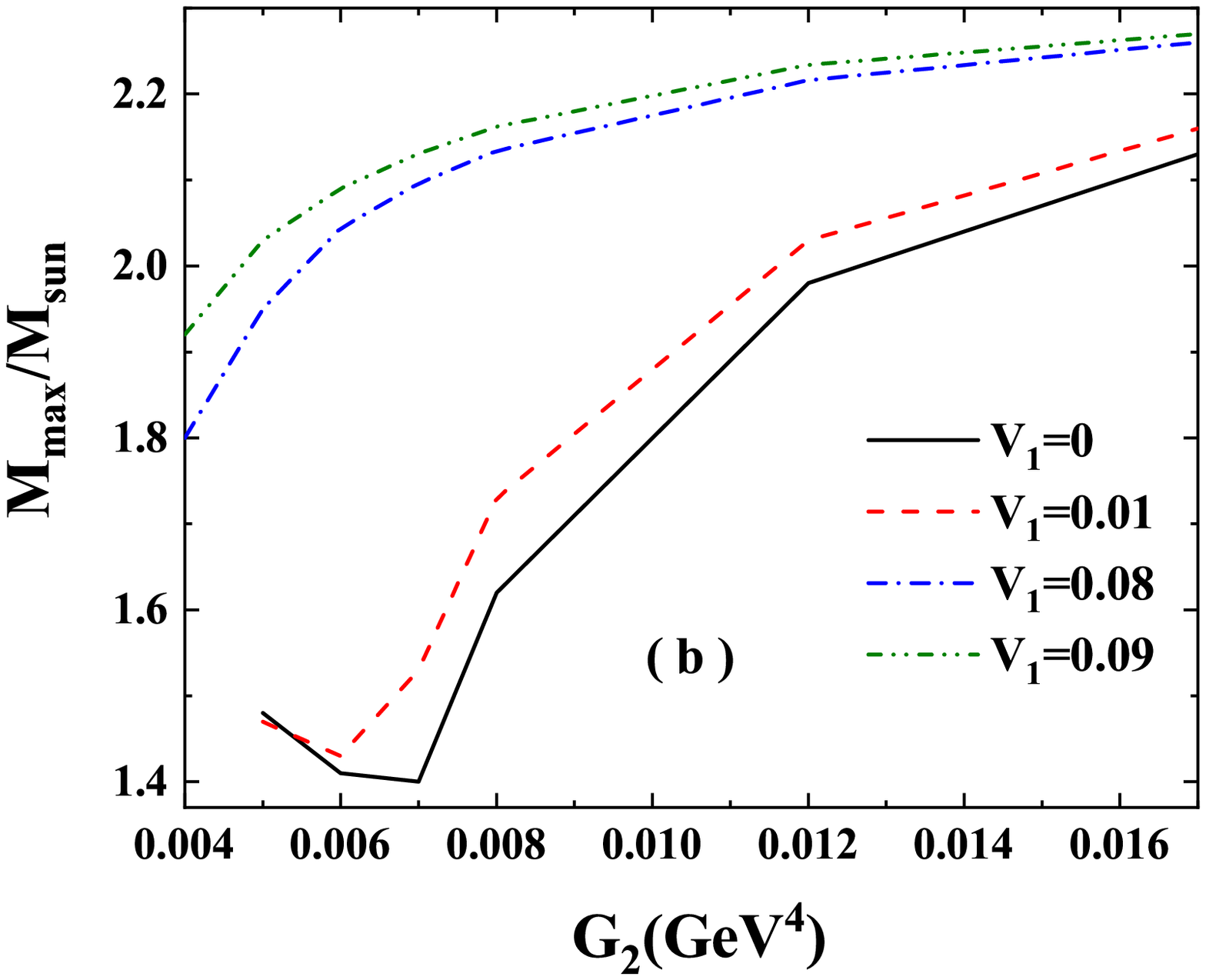}}
	\resizebox{0.5\textwidth}{!}{\includegraphics{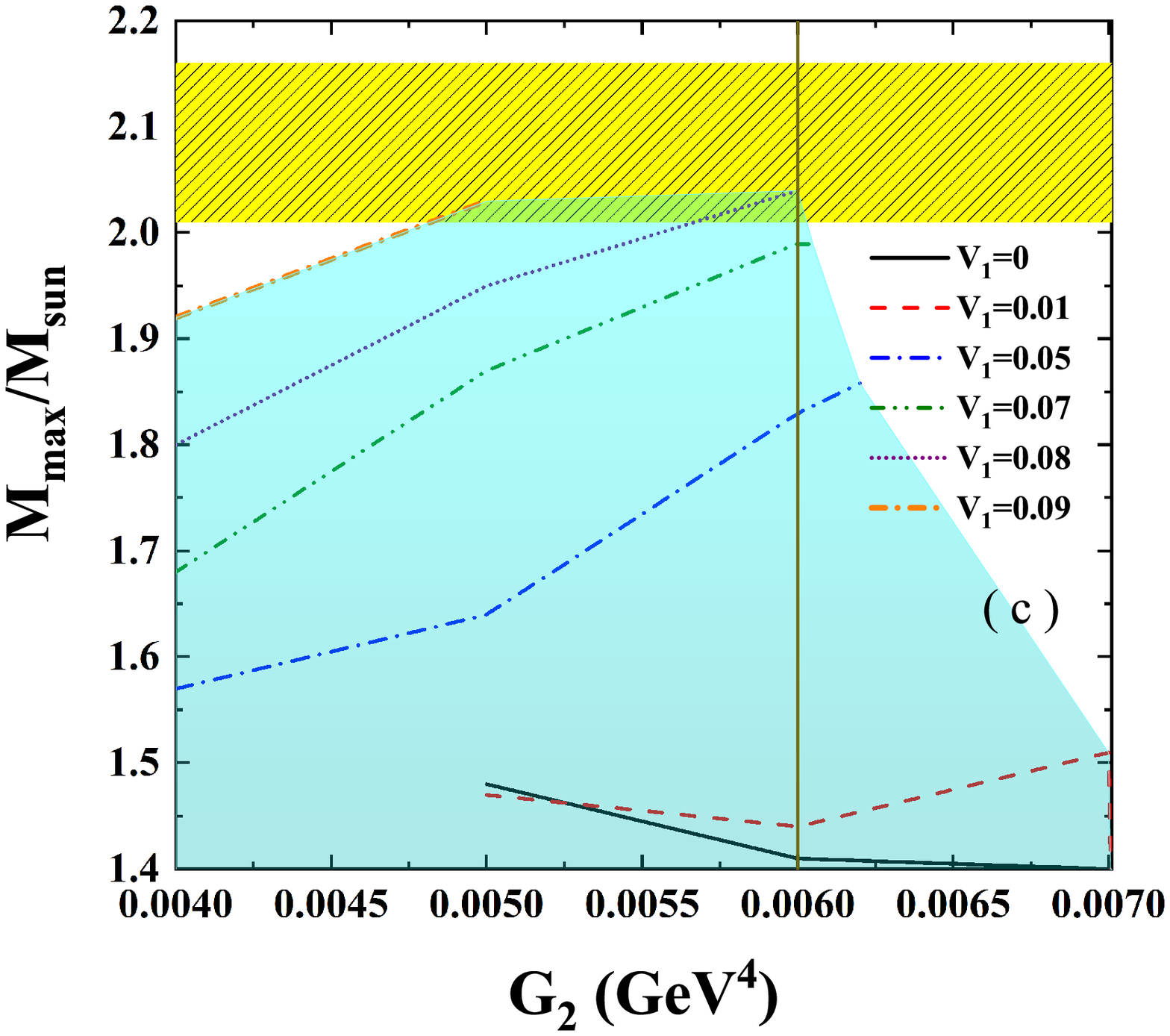}}
	\begin{center}
		\caption{{\small Panel (a),(b) : Maximum mass of the HS vs (a) $ q\bar{q} $ potential, $ V_{1} $ (GeV) (b) gluon condensate, $ G_{2} $ (GeV$ ^{4} $). 
			(c) : Maximum mass of "stable" HS vs gluon condensate, $ G_{2}  $ (GeV$ ^{4} $), for several values of $ q\bar{q} $ potential, $ V_{1}  $ (GeV). The dashed yellow region shows the constraint on the maximum mass of NSs, the shadowed blue region displays the values of FCM parameters for which a stable HS is predicted, and the vertical line manifests the adapted value of the gluon condensate which gives the critical temperature of about $ T_{c}=170 $ MeV}. \label{fig5}} 
	\end{center}
\end{figure*}

We also display the dependence of the maximum mass of the HS as a function of $q\bar{q} $ potential, $ V_{1} $, and gluon condensates, $ G_{2} $, in Fig.~\ref{fig5} (a) and \ref{fig5} (b), respectively. The  increasing  behavior of the maximum mass of the HS when increasing both the $q\bar{q} $ potential, $ V_{1} $, and the gluon condensate, $ G_{2} $ is obvious in Fig.~\ref{fig5}. In Fig.~\ref{fig5} (c), we display the maximum mass of a "stable" HS vs the gluon condensate, $ G_{2} $ for several values of $q\bar{q} $ potential, $ V_{1} $ (GeV). We also show the maximum mass constraint for NSs by the dashed yellow region and the values of FCM parameters through which stable HS is predicted by the shadowed blue area. The adapted value of $ G_{2} $ from lattice QCD which gives the transition temperature of about $ T_{c}=170 $ MeV is displayed by a vertical line. As is clear, the area in which all three constraints are satisfied occurs with a value of $q\bar{q} $ potential of about $ V_{1}=0.08 $ GeV.

Up to now, we have studied the effect of two FCM parameters, $q\bar{q} $ potential, $ V_{1} $, and and gluon condensates, $ G_{2} $, on the maximum mass of the HS. The calculation predicts that values of $ V_{1} $ as small as 0.01 GeV are excluded, since the maximum mass of stable HSs (around $ 1.4M_{\odot} $) is so far from the observational values. 
If one requires a stable HS, our calculations also exclude large values of $ G_{2} $, larger than around 0.007 GeV$^{4} $; and the maximum mass is shifted to values higher than $ 2M_{\odot} $, which is compatible with the observations. A stable HS is predicted for lower values of the gluon condensate, $ G_{2} $, around 0.006 GeV$^{4} $. Therefore, our calculations put constraints on the $q\bar{q} $ potential, $ V_{1} $, of around 0.08 GeV, and on the gluon condensates $ G_{2} $, of around 0.006 GeV$^{4} $. The adapted value of $ G_{2} $ from lattice QCD calculations which give rise to a critical temperature around $ T_{c}\sim170 $ is $0.006 $ GeV$^{4} $.

\subsection{Tidal deformability }\label{IV-C} 

Until 17 August 2017, electromagnetic observation of NSs~\cite{Lattimer:2006xb,Read:2008iy} and simultaneous measurements of both masses and radii of NSs~\cite{Ozel:2006km,Leahy:2007fb,Leahy:2008cq} provided constraints on the EOSs of such dence systems. However, these measurements are dependent on detailed modeling of the radiation and absorption mechanism at the NS surface and interstellar medium and are also subject to systematic uncertainties~\cite{Hinderer:2009ca}. Another possibility for obtaining information on the EOS of the NS is from ispiraling binary NSs due to the gravitational radiation. The tidal distortion of NSs in a binary system links the EOS describing NS matter, to the emission of the gravitational wave during the inspiral~\cite{Hinderer:2009ca}. 

On 17  August 2017, the first direct detection of a binary NS merger (GW170817) by the LIGO-Virgo scientific collaboration has opened a new window into modern astronomy. This historic detection has been instrumental in providing initial constraints on the tidal polarizibility (or deformability) of NSs~\cite{Negele:1971vb,Drischler:2016djf,Annala:2017llu,Drischler:2017wtt}. 

During the early regime of the inspiral, the signal is very clean, and the influence of the tidal effects is only a small correction to the wave form 's phase~\cite{Hinderer:2007mb}. The influence of the internal structure on the gravitational wave phase in this early regime of the inspiral is characterized by a single dimensionless parameter, namely, the ratio of the induced quadrupole moment to the perturbing tidal field (from the companion star). This ratio, which is called the tidal deformability,  (or tidal polarizability), $ \Lambda $, is related to the star's tidal Love number, $ k_{2} $, by
\begin{equation}\label{landa}
  \Lambda=\frac{2}{3}k_{2}(\frac{c^{2}R}{GM})^{5},
\end{equation} 
  where $  R $ and  $ M $ are the radius and mass of the NS. In other words, the tidal deformability $ \Lambda $ measures the star's quadrupole deformation in response to the companion's perturbing tidal field. The compactness of the star, $ C $, is defined as $ C=\dfrac{GM}{c^{2}R} $. As is clear,  $ \Lambda $ is extremely sensitive to the compactness of the star.

 the Tidal Love number $ k_{2} $, being a dimensionless parameter that is sensitive to the entire EOS~\cite{Hinderer:2007mb,Hinderer:2009ca}, is expressed as 
\begin{eqnarray}\label{k2}
&&k_{2}(C,y_{R})=\dfrac{8}{5}C^{5}(1-2C)^{2}[(2-y_{R})+2C(y_{R}-1)]
\nonumber\\
&&\times\{2C(6-3y_{R}+3C(5y_{R}-8))
\nonumber\\
&&+4C^{3}[(13-11y_{R})+C(3y_{R}-2)+2C^{2}(1+y_{R})]
\nonumber\\
&&+3(1-2C)^{2}[(2-y_{R})+2C(y_{R}-1)]\log(1-2C)\}^{-1}~.\nonumber\\
\end{eqnarray}

Now we proceed to compute $ y_{R} $, which is the value of the function $ y(r) $ at the surface of the star (for more details, see Refs.~\cite{Hinderer:2007mb,Bildsten:1992my,Flanagan:2007ix,Hinderer:2009ca,Postnikov:2010yn,Piekarewicz:2018sgy} and references contained therein). $ y(r) $ satisfies the following nonlinear, first-order differential equation~\cite{Postnikov:2010yn,Fattoyev:2012uu}:

\begin{eqnarray}\label{dy}
&&r\frac{dy(r)}{dr}+y^{2}(r)+F(r)y(r)+r^{2}Q(r)=0~;
\nonumber\\
&&\text{with}~y(0)=2 ~~~\text{and}~~~ y_{R}=y(r=R)\nonumber\\
\end{eqnarray}

where $ F(r) $ and $ Q(r) $ are functions of the mass, pressure,  and energy density profiles assumed to have been obtained by solving the TOV equations and are given by the expressions
\begin{equation}\label{f}
F(r)=\dfrac{1-4\pi Gr^{2}(\epsilon(r)-P(r))}{(1-\dfrac{2G M(r)}{r})}
\end{equation}
and
\begin{eqnarray}\label{q}
&&Q(r)=
\nonumber\\
&&\dfrac{4\pi}{(1-\dfrac{2G M(r)}{r})}(5\epsilon(r)+9P(r)+\dfrac{\epsilon(r)+P(r)}{c_{s}^{2}(r)}-\dfrac{6}{4\pi r^{2}})
\nonumber\\
&&-4[\dfrac{G(M(r)+4\pi r^{3}P(r))}{r^{2}(1-\dfrac{2GM(r)}{r})}]
\end{eqnarray}
in which $c_{s}^{2}(r)=dP(r)/d\epsilon(r)  $ is the speed of sound at radius $ r $.

One may use the weighted $ \tilde{\Lambda}({\cal M})$ where ${\cal M}  $ is defined by $ {\cal M}=M_{1}^{3/5}M_{2}^{3/5}/(M_{1}+M_{2})^{1/5} $. However, as both EOSs of the NSs are the same, the mass ratio of stars has no big effect on $ \tilde{\Lambda}$. Therefore we can use $ \Lambda $ instead of $ \tilde{\Lambda}$ without loss of generality ~\cite{Postnikov:2010yn}.

 GW170817 puts only an upper limit on the tidal deformability of a $ 1.4 M_{\odot} $ NS, i.e., $ {\Lambda}_{1.4}\leq800 $~\cite{Piekarewicz:2018sgy}. Moreover, the authors in Ref~\cite{Most:2018hfd} find  additional constraints on the tidal deformability and radii of neutron and hybrid stars. For a purely hadronic star with a mass of $ 1.4 M_{\odot} $, the radius of the NS is considered to be $ 12.00 $ km $<R_{1.4}$ $<13.45 $ km; similarly, the smallest weighted average dimensionless tidal deformability is $ \tilde{\Lambda}_{1.4}>375 $. Since EOSs with a phase transition allow for very compact stars on the "twin star" branch, small radii are possible for HSs~\cite{Montana:2018bkb}; therefore,  the radius varies in a much broader range of $ 8.35$ km $<R_{1.4}$  $ <13.74  $ km, with $ \tilde{\Lambda}_{1.4}>35.5 $. 

In order to check these new constraints, we have computed the tidal deformabililty for individual stars with the mass of 1.4 $ M_{\odot} $: $ \Lambda_{1.4} $~\cite{Hinderer:2009ca,Postnikov:2010yn}. The results of the computation of $ y_{R} $, the compactness $ C $, tidal Love number $ k_{2} $, and dimensionless tidal deformability $ \Lambda $ for HSs with the mass of 1.4 $ M_{\odot} $ within the FCM, with several choices of parameter sets, are summarized in Tables~\ref{t3} and~\ref{t4} . Table~\ref{t3} concerns very low values of the $ q\bar{q} $ potential, $ V_{1}=0, 0.01 $, with several choices of gluon condensate $ G_{2} $, while Table~\ref{t4} collects the results of higher values of $ V_{1}=0.05,0.07,0.08,0.09,0.1 $ GeV.

As seen in Table~\ref{t3} and~\ref{t4}, for the cases in which the mass of $ 1.4M_{\odot} $ occurs on the hadron branch, the mentioned properties are similar for the same hadron interaction. The reason is that the EOS of hadron matter governs the star in the hadron branch and as is clear from Eqs.~(\ref{dy}), ~(\ref{f}) and ~(\ref{q}), $ y_{R} $, which depends on the profile of the star, takes the same value, so $ k_{2} $ and hence $ \Lambda $, from Eqs~(\ref{k2}) and (\ref{landa}) will have a unique value for the same hadron interaction. In these cases, the HSs become much less compact, and tidal defromability takes larger values in comparison with the cases in which the mass of $ 1.4M_{\odot} $ occur on the quark branch. In those cases, the EOS is the hybrid EOS of hadron and quark matter within the Maxwell construction. If one compares these results with those for pure NSs~\cite{Khanmohamadi:2019jky}, one can see that the HSs are a little less compact in comparison with pure NSs. Moreover, if we compare the result for HSs within the FCM with HSs within the MIT and NJL models~\cite{Khanmohamadi:2019jky}, it is obvious that when a star with the mass of $ 1.4M_{\odot} $ occurs in the hadron branch, the tidal deformability parameters are independent of the employed quark models and just depend on hadron models. It means that the tidal deformability in such cases is the same for a specified hadron model with any quark models.

For very low values of $ q\bar{q} $ potential, $ V_{1} =0, 0.01$ GeV, and low values of quark condensation, $ G_{2}=0.005, 0.006 $ GeV$ ^{4} $ and rarely $ G_{2}=0.007 $ GeV$ ^{4} $, the star mass of $ 1.4M_{\odot} $ occurs on the quark branch. In such cases, the HSs are much compacted ( $ 9$ km $<R_{1.4}<11.6 $ km) and the dimensionless tidal deformability takes low values (lower than around $ \Lambda_{1.4}=350 $).

For larger values of the $ q\bar{q} $ potential, $ V_{1} \geq 0.05$ GeV, and all values of quark condensate, $ G_{2} $, (except in the case of $ V_{1} =0.05$ GeV and $ G_{2} =0.004$ GeV$ ^{4} $ ), the mass of $ 1.4M_{\odot} $ occurs on the hadron branch, and therefore the tidal defrormability depends only on hadron interaction. The values of dimentionless tidal defromability are in the range $ 470<\Lambda_{1.4}<485 $, and the radii of the HSs in such cases are in the range $12.28$ km $ <R_{1.4} <12.42$ km. All the results are in line with the constraint on tidal deformability for HSs, $ 35.5<\tilde{\Lambda}_{1.4}<800 $.

If we link the results of this section for tidal deformability to the results of the last section on the maximum mass of the HSs, we can observe that for the cases with larger masses  (which are compatible with observations), the tidal defrormability takes larger values, which is more compatible with the constraints extracted from binary  GW170817 for "NSs": $ 375<\tilde{\Lambda}_{1.4}<800 $. Therefore, this scenario is a feasible scenario for a NS.

\begin{figure*}[htb]
	\vspace{-0.50cm}
	\resizebox{0.48\textwidth}{!}{\includegraphics{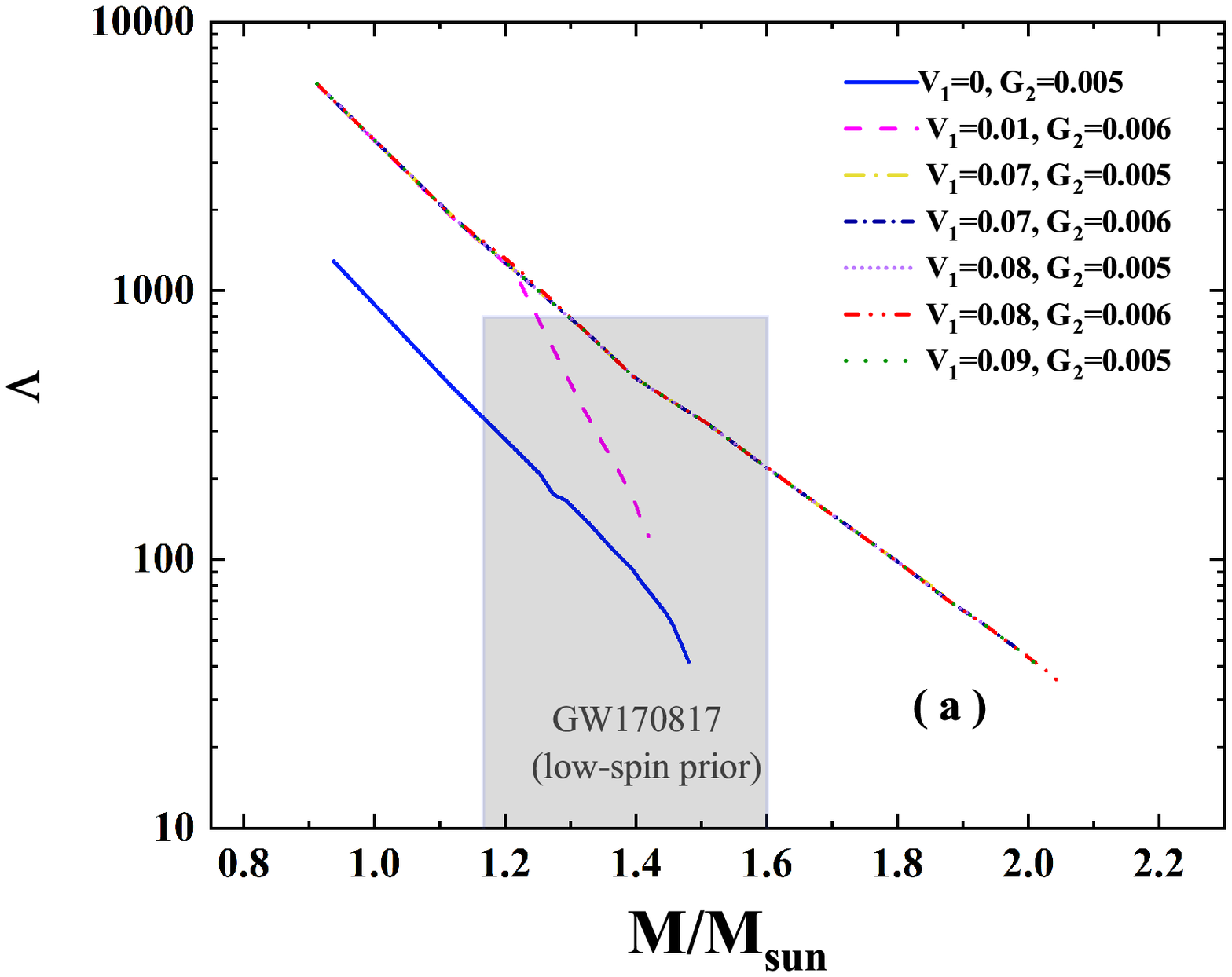}}
	\resizebox{0.48\textwidth}{!}{\includegraphics{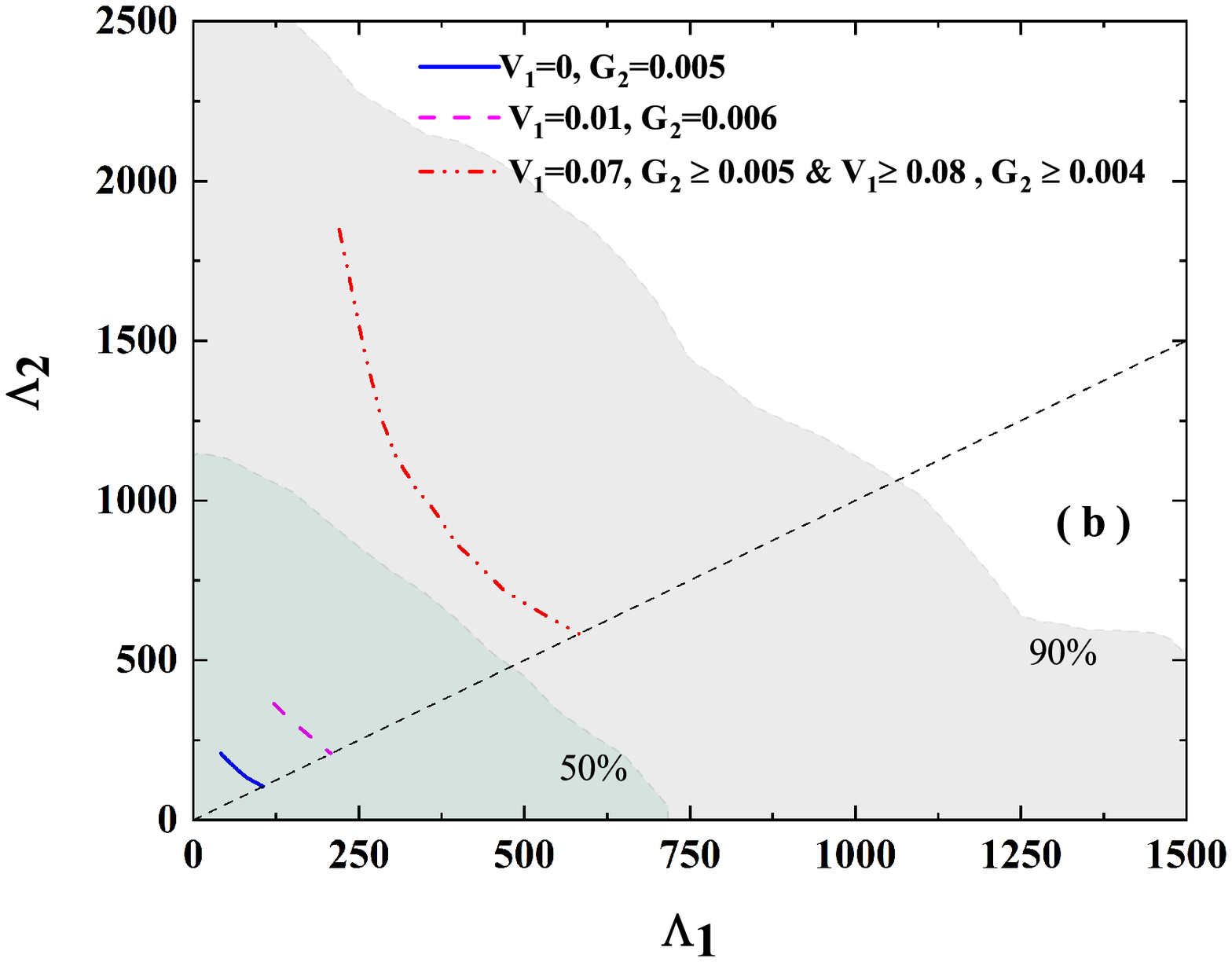}}

	\begin{center}
		\caption{{\small (a)  Dimentionless tidal deformability $ \Lambda $ as a function of the star mass, $\text{M}/M_{\text{sun}}$, for some of the parameter sets concerning stable hybrid stars mentioned in Fig.~\ref{fig5} (c). The gray box shows the $ \Lambda \leq800$ constraint in the range of $ 1.16M_{\odot}-1.60 M_{\odot} $ for the low-spin prior~\cite{TheLIGOScientific:2017qsa,Marczenko:2018jui}. (b) Corresponding tidal deformability $ \Lambda_{1} $ and $ \Lambda_{2} $ of the low- and high-mass mergers obtained from the $ \Lambda(m) $. The 50 $ \% $ and 90$ \% $ fidelity regions of the low-spin prior are also shown~\cite{TheLIGOScientific:2017qsa,Marczenko:2018jui}. 	}} \label{fig6} 
	\end{center}
\end{figure*}

Moreover, we study the effect of  different chirp masses and different binary mass ratios $ q=M_{2}/M_{1} $ on the tidal deformability $ \Lambda $. The total mass of $ M_{\text{tot}}=M_{1}+M_{2}\simeq2.74M_{\odot} $, which was inferred from the gravitational wave signal, is compatible with masses measured in binary NS systems containing pulsars~\cite{Ozel:2016oaf,Bauswein:2019skm}. The binary mass ratio $ q $ is restricted to the range  0.7 to 1. In Fig.~\ref{fig6} (a), we present the tidal deformability $ \Lambda $ as a function of star mass $ \text{M}/M_{\text{sun}} $. The gray box shows the $ \Lambda\leq800 $ constraint in the range of $ 1.16M_{\odot}-1.60M_{\odot} $ of the low spin prior~\cite{TheLIGOScientific:2017qsa,Marczenko:2018jui}. As seen in Fig.~\ref{fig6} (a), the hybrid EOSs mentioned in Fig.~\ref{fig5} (c), associated with stable hybrid stars, are within the range of this constraint. In Fig.~\ref{fig6} (b), we display the tidal deformability $ \Lambda_{1} $ and $ \Lambda_{2} $ of the low- and high-mass mergers obtained from the $ \Lambda(m) $. For comparison, the  $ 50\% $ and $90\% $ probability contours of the low-spin prior from the analysis by the LIGO VIRGO Collaboration (LVC) of the gravitational wave signal of the GW170817 merger event are also shown ~\cite{TheLIGOScientific:2017qsa,Marczenko:2018jui}. As seen in Fig.~\ref{fig6} (b), the hybrid EOSs with very low values of $ q\bar{q} $ potential, $ V_{1}= 0, 0.01$ GeV, are in the range of $ 50\% $ fidelity region and the hybrid EOSs with high values of $ V_{1}>0.05$ GeV are in $ 90\% $ fidelity region.

\begin{figure*}[htb]
	\vspace{-0.50cm}

	\resizebox{0.6\textwidth}{!}{\includegraphics{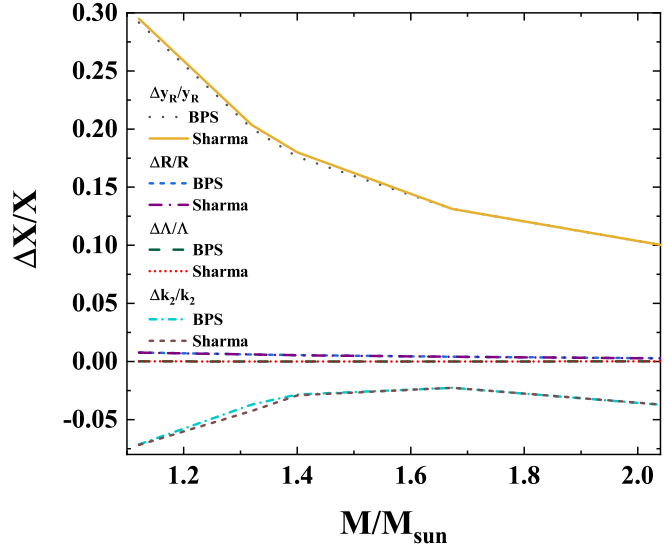}}
	\begin{center}
		\caption{{\small  Relative deviation $(X_{ \text{BPS or Sharma} }-X_{\text{HW}})/X_{\text{HW}} $
				for the different quantities radius R, tidal Love number $ k_{2} $, dimensionless tidal deformability $ \Lambda $, and $ y_{R} $  calculated with  BPS and Sharma crust and the quantities  calculated with HW crust, as a function of hybrid star mass $\text{M}/M_{\text{sun}}$ for the parameter set $ V_{1}=0.08 $ GeV and $ G_{2} =0.006$ GeV$ ^{4} $ of the FCM. See
				text for details.}\label{fig7} } 
	\end{center}
\end{figure*}

We also study the influence of different inner and outer crusts on the radius and tidal deformability of the hybrid stars. We apply two different crustal EOSs: the first one is that of Bame, Pethick, and Suttherland (BPS)~\cite{Baym:1971pw}, and the second one uses the base of microscopic calculation (we mentioned it as "Sharma" in the figures)~\cite{Sharma:2015bna}. In Fig.~\ref{fig7}, we display the relative deviation, for different quantities namely radius R, tidal Love number $ k_{2} $, dimensionless tidal deformability $ \Lambda $ and $ y_{R} $, calculated with  BPS and Sharma crusts and the quantities  calculated with the HW crust,  $ \dfrac{X_{ \text{BPS or Sharma} }-X_{\text{HW}}}{X_{\text{HW}}} $ as a function of the star mass $\text{M}/M_{\text{sun}}$. We present the calculation for a sample parameter set $ V_{1}=0.08 $ GeV with $ G_{2}=0.006 $ GeV$ ^{4} $ of the FCM. As is clear from the Figures, the EOSs of the crusts are more or less important in the determination of all these quantities except for the dimensionless tidal deformability $ \Lambda $. This result arises from a cancellation between the second Love number $ k_{2} $ and the stellar compactness $ C $. Whereas $ \Lambda $ depends on both $ C $ and $ k_{2} $, with $ k_{2} $ being a highly complex function of $ C $ and $ y_{R} $ [see eq.~\ref{k2}], the value $  \Lambda\propto k_{2}C^{-5} $ is almost equal for different crusts. So, while $ y_{R} $ and hence the second Love number $ k_{2} $ is sensitive to the crustal component of the EOS, such sensitivity disappears in the case of the dimensionless tidal deformability $ \Lambda $. As shown earlier, the behavior of $ \Lambda $ is largely dictated by the EOS of the uniform liquid core.  It worth noting that the core-crust transition densities in BPS and Sharma crustal EOSs are almost the same (around 0.06fm$ ^{-3} $), and the results obtained on the HS radii when applying them are almost the same. Meanwhile, the core-crust transition density in HW crustal EOSs ( 0.04 fm$ ^{-3} $) is different and the results in this case are a little bit different (in HW crustal EOSs the radii of the stars are around 0.04-0.1 km lower than the mentioned cases). It seems that - at least in the cases studied - the core-crust transition density has more influence in the HS radii than the type of the crustal EOS. Our results are in good agreement with those given in Refs~\cite{Piekarewicz:2018sgy,Perot:2020gux}. 

We close this subsection by checking the constraints deduced from GW170817 for the radius of NS; i.e., $ R_{\text{max}}>9.6 $ km and  $ R_{1.6}>10.7 $ km~\cite{Marczenko:2018jui,Bauswein:2019skm}. 

\begin{figure*}[htb]
	\vspace{-0.50cm}
	\resizebox{0.6\textwidth}{!}{\includegraphics{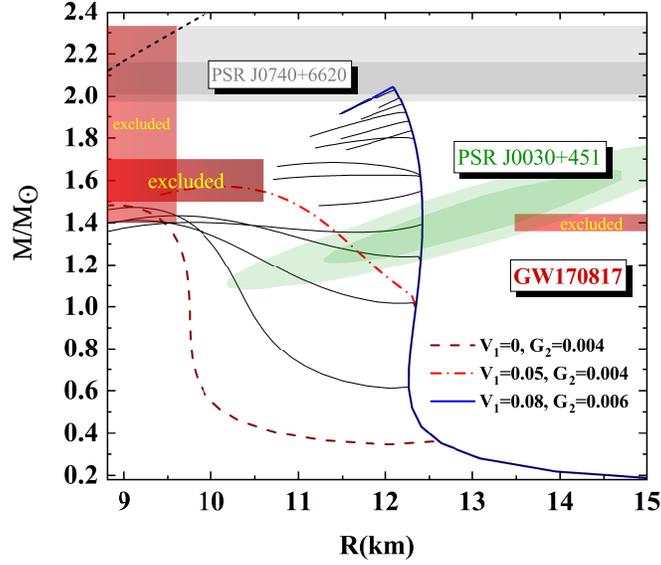}}

	\begin{center}
		\caption{{\small Mass-radius relations of the stable hybrid star mentioned in Fig.~\ref{fig5} (c) overplotted with constraints on the NS radii and maximum mass. The gray region shows the constraint on the maximum mass extracted from the PSR J0740+6620 results~\cite{Cromartie:2019kug}.
		The red region shows the excluded region of mass-radius relation inferred from the GW170817 results~\cite{Bauswein:2019skm,Marczenko:2018jui}. The green region shows the constraint on mass-radius relation extracted from NICER, PSR J0030+451~\cite{Miller:2019cac,Marczenko:2020jma}.}
		} \label{fig8} 
	\end{center}
\end{figure*}

 We also check the new constraint on mass-radius relation extracted from the neutron star interior composition explorer (NICER) for PSR J0030+451~\cite{Miller:2019cac}, as well as the constraint on the maximum mass extracted from PSR J0740+6620~\cite{Cromartie:2019kug}. These constraints are summarized in Fig.~\ref{fig8}. In this figure, we have also shown various stable hybrid stars' mass-radius relations from Fig.~\ref{fig5} (c). The green (red) region shows the constraint on the mass-radius relation inferred from NICER for PSR J0030+451 (the excluded region inferred from the binary GW170817). The constraint on maximum mass, extracted from PSR J0740+6620, is shown by the gray region. Finally, the dashed line shows the causality constraint. 
  As is clear from the figure, the hybrid stars with very low values of $ q\bar{q} $ potential, $ V_{1}\leq0.01 $ do not fulfill the value and the radius of maximum mass, while they fulfill the constraints on $ R_{1.6} $ and $ R_\text{{max}} $. These cases also satisfy the constraint inferred from NICER. The only exception is the case with $ V_{1}=0 $ and $ G_{2}=0.004 $. 
   The constraint on $ R_{1.6} $ is not fulfilled only in the case with $ V_{1}=0.05 $ and $ G_{2}=0.004 $.
  
   The hybrid stars with higher values of  $q\bar{q} $ potential, $V_{1}\geq0.07 $, and gluon condensate, $ G_{2}>0.05 $ fulfill the maximum mass constraint from PSR J0740+6620 and the mass-radius constraints inferred from both GW170817 and NICER, PSR J0030+451.

\section{CONCLUSION}\label{V}

In this paper, we have studied the appearance of a quark matter in the NS core with the corresponding quark-gluon EOS derived in the framework of the FCM. We performed our analysis at various constant values of the parameters of the model namely, the gluon condensate, $ G_{2} $, and the $ q\bar{q} $ potential, $ V_{1} $, extracted from QCD sum rules and lattice data, respectively. For small values of $ q\bar{q} $ potential, $ V_{1}\leq0.01 $, the HS with maximum mass higher than $ 2M_{\odot} $ is predicted for large values of gluon condensate, $ G_{2} $, around 0.012 GeV$ ^{4} $; however, these HSs become unstable as soon as the onset of the quark phase in the core of the star. This instability manifests itself as a cusp in the mass-radius curves. The large discontinuity in the energy density  is probably responsible for the instability of the quark core, since the star cannot counteract the additional pressure due to the additional force exerted on the star. 

By increasing the values of $ q\bar{q} $ potential, $ V_{1} $, to higher than 0.07 GeV, a stable HS with maximum mass higher than $ 2M_{\odot} $ is predicted also for small values of the gluon condensate, $ G_{2} $, around 0.006 GeV$ ^{4} $. A stable HS with a maximum mass of  $ 2.03M_{\odot} $  ($ 2.04M_{\odot} $) is calculated for $ G_{2}=0.005 $ GeV$ ^{4} $ and $ V_{1}= 0.09$ GeV ($ G_{2}= 0.006$ GeV$ ^{4} $ and $ V_{1}=0.08 $ GeV). 

Strictly speaking, our calculations excluded very low values of  $ q\bar{q} $ potential $ V_{1} $, since the maximum mass of stable HSs for very low values of $ V_{1} $ is around $ 1.4M_{\odot} $, which is so far from the observational values. Besides, it suggested that values of $ q\bar{q} $ potential, $ V_{1} $, around 0.08-0.09 GeV and gluon condensate, $ G_{2} $, around 0.006 GeV$^{4} $ are the FCM parameters in which stable HSs with maximum mass higher than $ 2M_{\odot} $ are predicted. Since the lattice calculations predict the value of the gluon condensate to be $ G_{2}=0.006 $ GeV$^{4} $, resulting in a critical temperature of about $ T_{c}=170 $ MeV, in the parameter set $ G_{2}= 0.006$ GeV$^{4} $ and $ V_{1}=0.08 $, we obtain a maximum mass vale for a stable HS of $ 2.04M_{\odot} $. Therefore, one can conclude that $ V_{1}=0.08 $ GeV could be the best choice in accordance with our calculation.

In order to test the new constraint which was extracted from the gravitational waves of the binary GW170817 on tidal deformability and hence on the radii of a NSs, we have calculated the tidal deformability of a HS with the mass of $ 1.4M_{\odot} $ with several choices of parameter sets of the FCM. For very low values of quark-antiquark static potential, $ V_{1}=0,0.01 $ GeV, and values of gluon condensate $ G_{2} $ lower than around 0.007 GeV$ ^{4} $, the mass of $ 1.4M_{\odot} $, occur on the quark branch, and so the HS becomes so compact ( $ 9$ km $<R_{1.4}<11.6 $ km) then the tidal deformability takes lower values ($ 84<\Lambda_{1.4}<345$) . However, even in such cases, the value of tidal deformability and hence the radii of HSs is still compatible with the constraints in HSs, $ \tilde{\Lambda}_{1.4}>35.5 $ and $ 8.35 $ km $<R_{1.4}$   $ <13.74  $ km and. The lower limit of the constraint in the HS is much lower than that for purely NSs because of the probability of the existence of a "twin branch" in HSs.

For higher values of quark-antiquark potential, $ V_{1} \geq0.05$, the mass of $ 1.4M_{\odot} $ occurs on the hadron branch. Thus the HS becomes much less compact, $12.28$ km $ <R_{1.4} <12.42$ km, and therefore the tidal deformability takes larger values, $ 470<\Lambda_{1.4}<485 $, for different hadron interactions supplemented by TBF. These values are more compatible with the constraint for tidal deformability for NSs that is, $ 375<\tilde{\Lambda}_{1.4}<800 $ and  $ 12.00 $ km $<R_{1.4}$ $<13.45 $ km. In such cases the value of tidal deformability is independent of the quark model and only depends on the hadron model and the hadron interaction. As we mentioned above, in some of such cases our calculations predict a stable HS with a pure quark core with the maximum mass higher than $ 2M_{\odot} $, which is in the recent constraint put on maximum mass that is,  $ 2.01^{+0.04}_{-0.04}\leq{M}_{\text{max}}/{M\odot}\lesssim 2.16^{+0.17}_{-0.15}$.

We study the effect of different chirp masses and different binary mass ratios $ q=M_{2}/M_{1} $ on the tidal deformability $ \Lambda $. All the hybrid EOSs are in the range of constraint on the low-spin prior.

The influence of different inner and outer crusts on the tidal deformability of the stars is examined. The crustal EOS is important in the determination of the radius R, $ y_{R} $ and the tidal Love number $ k_{2} $,  but not the dimensionless tidal deformability $ \Lambda $. This result arises from a cancellation between the second Love number $ k_{2} $ and the stellar compactness $ C $.

We check the new constraint extracted from NICER for PSR J0030+451. All the hybrid EOSs except the case with $ V_{1}=0 $ and $ G_{2}=0.004 $ are within the range of this constraint. We also check the constraints on the radius of maximum mass and the star with $ 1.6M_{\odot} $ configurations extracted from GW170817. The hybrid EOSs with $ V_{1}>0.05 $ satisfy these constraints. 

 Considering all the above results, we conclude that, in some range of the parameter sets of the FCM, i.e., $ V_{1}\sim0.08-0.09$ GeV, and the gluon condensate, $ G_{2}\sim0.005-0.006 $ GeV$ ^{4} $, we find stable HSs with a maximum mass higher than $ 2M_{\odot} $ in which the tidal deformability of a HS is exactly compatible with the new constraint extracted from the binary GW170817 for NSs. Also, it is compatible with the mass-radius constraints extracted from both GW170817 and PSR J0030+451. Therefore, this scenario for the EOS of the NS system can be considered an acceptable one.

\section*{Acknowledgments}
We warmly appreciate G. F. Burgio from INFN Catania turning our attention to the FCM model, and we would also like to thank the Research Council of the University of Tehran. S. A. T. is grateful to the School of Particles and Accelerators, Institute for Research in Fundamental Sciences.


%
%

%

\begin{table*}[htb]
	\begin{tabular}{cccccccccccc}
		\hline\hline
	
		$%
		V_{1} $  & 	$ G_{2} $ & $\mu _{B}$ &   $ {\rho_{B}^{(1)}}/{\rho_{0}}$ & $ {\rho_{B}^{(2)}}/{\rho_{0}}$ & $\epsilon ^{(1)}$ & $
		\epsilon ^{(2)}$ & $ \rho_{CB\text{max}}/{\rho_{0}} $ $  $ & ${\text{R}}_\text{max}$ & ${\text{M}_{\text{max}}}({\text{M}_{\odot}})$   \\ \hline

	0  & 0.005 & 987.3 & 1.25 &2.66 & 191.8 &415.2 &10.10 &8.89 & 1.48\\ 
		& 0.006 & 1072.1 & 2.09 &3.46  & 330.2& 565.7 &10.52 &9.16 &1.41\\ 
	&0.007 & 1146.6 & 2.57 & 4.28 & 416.8 & 729.9 &9.5 &9.79 &1.40\\ 
		&0.008 & 1206.2 & 2.9 & 5.02 & 478.1 &887.8 & 5.06&12.39 &1.62\\ 
	&	0.012 & 1364.5 & 3.62 & 7.38 & 626.9 & 1448.1 & 7.47&  12.15 & 1.98\\
	 &0.017 &1488.3 & 4.10 & 9.66 & 737.0 & 2061.4 & 9.7 &  11.93 & 2.13\\
		\hline
	 0.01&0.005 &1018.7 & 1.63 & 2.81 & 253.1 &444.9  & 10.0& 9.11&1.47\\ 
	&	0.006 &1111.0 & 2.36 &3.71  & 377.2& 618.1 &9.75 &9.58 &1.43\\ 
		&0.007 & 1184.3 & 2.78 & 4.55 & 455.8 & 791.3 &4.62 &12.39 &1.53\\ 
		 &0.008 & 1241.6 & 3.07 & 5.29 & 512.2 &953.2 & 5.40&12.34 &1.73\\ 
	 &0.012 & 1393.8 & 3.74 & 7.62 & 653.1 & 1519.9 & 7.68&12.10 &2.03\\
			& 0.017 & 1514.3 & 4.2 & 9.88 & 759.9 & 2138.4 & 10.0&11.87 &2.16\\
		
		\hline\hline

	\end{tabular}
	\caption{{\small Hadron-quark phase transition and hybrid star structure properties for several values of gluon condensate, $ G_{2} $  (GeV$ ^{4} $) and $ q\bar{q} $ potential, $ V_{1} $  (GeV). where $ \mu_{B} $ is critical baryon chemical potential (MeV),  $ {\rho_{B}}/{\rho_{0}}$ is the ratio of the baryon density  to the saturation density and  $ \epsilon $ is the energy density  at the starting $ (1) $ and ending point $ (2) $ of phase transition (MeV/fm$ ^{3} $),   $ {\text{M}_{\text{max}}}({\text{M}_{\odot}}) $ is the Maximum mass of the star in terms of the sun mass, $\rho_{CB\text{max}}/{\rho_{0}}   $ is the ratio of central density to the saturation density,  and  $ {\text{R}}_\text{max}$ is the hybrid star's radius (km) }}\label{t1}
\end{table*}

\begin{table*}[htb]
	\begin{tabular}{cccccccccccc}
		\hline\hline
	
		$%
		V_{1} $  & 	$ G_{2} $ & $\mu _{B}$ &   $ {\rho_{B}^{(1)}}/{\rho_{0}}$ & $ {\rho_{B}^{(2)}}/{\rho_{0}}$ & $\epsilon ^{(1)}$ & $
		\epsilon ^{(2)}$ & $ \rho_{CB\text{max}}/{\rho_{0}} $ $  $ & ${\text{R}}_\text{max}$ & ${\text{M}_{\text{max}}}({\text{M}_{\odot}})$   \\ \hline

		0.05  & 0.004 & 1075.35 & 2.11 &2.77 & 334.4 &448.3 &8.26 &10.12 & 1.57\\ 
		 0.07&  & 1206.2 & 2.90 &3.77  & 478.1& 645.6 &6.12 &11.39 &1.68\\ 
		 0.08& & 1266.9 & 3.19 & 4.28 & 536.58 & 757.63 &4.81 &12.17 &1.8\\ 
		 0.09& & 1322.167 & 3.44 & 4.78 & 587.7 &870.5 & 5.0&12.2 &1.92\\ 
		 0.1& & 1372.9 & 3.65 & 5.24 & 634.4 & 983.1 & 5.31&  12.14 & 2.0\\
			0.12&  &1464.5 & 4.01 & 6.12 & 716.3 & 1208.3 & 6.18 &  11.97 & 2.11\\
		\hline
		 0.05&0.005 &1203.5 & 2.89 & 4.07 & 475.5 &703.8  & 5.27& 11.83&1.64\\ 
		 0.07& &1301.4 & 3.35 &4.92  & 568.7& 890.4 &5.0 &12.26 &1.87\\ 
		 0.08& & 1347.3 & 3.55 & 5.31 & 610.9 & 989.9 &5.43 &12.17 &1.95\\ 
		 0.09& & 1391.1 & 3.73& 5.70 & 650.6 &1089.9 & 5.83&12.1 &2.03\\ 
		 0.1& & 1432.9& 3.89 & 6.09 & 686.7 & 1192.3 & 6.18&12.02 &2.08\\
		0.12 	& & 1512.5 & 4.2 & 6.85 & 758.3 & 1403.4 & 7.0&11.87 &2.16\\
			\hline
		0.05&0.006 &1281.1 & 3.26 & 5.02 & 550.3 &911.04  & 5.14& 12.29&1.83\\ 
		0.07& &1363.5 & 3.62 &5.74  & 626.1& 1088.4 &5.83 &12.16 &1.99\\ 
		0.08& & 1403.3 & 3.78 &6.1  & 661.2& 1182.5  &6.187 &12.08 &2.04\\ 
		0.09& & 1441.8 & 3.92 & 6.45 & 695.2 &1278.6&6.58&12.0 &2.09\\ 
		0.1& & 1479.4 & 4.07 & 6.81 & 729.3 & 1377.2 & 6.81&11.95 &2.13\\
		0.12&  & 1551.6 & 4.33 & 7.5 & 792.5 & 1580.3 & 7.62&11.79 &2.19\\
		\hline\hline

	\end{tabular}
	\caption{{\small Same as Table~\ref{t1}  but for higher values of $ q\bar{q} $ potential, $ V_{1} $  (GeV) of the FCM}\label{t2}}
\end{table*}

\begin{table*}
	\begin{center}
		\begin{tabular}{ccccccccc}
			\hline\hline
		$ V_{1} $&~~~ $ G_{2} $& $ \rho_{CB} $  & ${\text{R}}$ & ${\text{C}}$ &$y_R$&$ k_{2} $&$\Lambda$\\ \hline

			0~~~& 0.005~~~&1.065 ~~~& 9.44  ~~~&0.219~~~&0.393~~~&0.0646~~~&84.573 \\			
		& 0.006 &1.42 & 9.47 & 0.218 &0.388 &0.0655 &88.396 \\
	 &0.007&1.39& 10.01 & 0.206&0.380 &0.0720 &128.498 \\ 
	 
		0.01 ~~~&0.005  &  1.078 & 9.72 & 0.212&0.388 &0.0685 & 105.545 \\ 
		&0.006  & 1.156 & 10.27 & 0.201& 0.378 & 0.0750   &152.721  \\
		& 0.007 & 0.42& 12.42 & 0.168& 0.356 & 0.0951 & 469.980 \\ 
	& $ \geq0.008 $ & 0.42& 12.42 & 0.168& 0.356 & 0.0951 & 469.980  \\

		\hline\hline
		\end{tabular}
		\caption{{\small  Central density $ \rho_{CB} $ (fm$ ^{-3}$), radius R(km), compactness $ C $, $ y_{R} $, tidal Love number $ k_{2} $ and dimensionless tidal deformability $ \Lambda $ for several hybrid stars with the mass of $ 1.4M_{\odot} $ studied in the paper. The units of $ V_{1} $ and $ G_{2} $ are GeV and GeV$ ^{4} $, respectively}.}\label{t3}
	\end{center}
\end{table*}

\begin{table*}
	\begin{center}
		\begin{tabular}{ccccccccc}
			\hline\hline
			$ V_{1} $~~~& $ G_{2} $~~~& $ \rho_{CB} $  & ${\text{R}}$ & ${\text{C}}$ &$y_R$&$ k_{2} $&$\Lambda$\\ \hline

			0.05~~~& 0.004~~~&0.72~~~ & 11.28~~~  &0.183~~~&0.365~~~&0.0861~~~&279.02 \\			
			 0.055~~~& 0.004 ~~~&0.65 ~~~& 11.75~~~ & 0.175~~~& 0.365 ~~~& 0.0906   ~~~&361.243  \\
			 
			 $ \geq0.05 $~~~& $ \geq0.005 $  ~~~&  0.42~~~ & 12.42~~~ & 0.168~~~& 0.355~~~ & 0.0958 ~~~  &469.980 \\
			 
			 $ \geq0.063 $~~~& $ \geq0.004 $  ~~~&  0.42~~~ & 12.42~~~ & 0.168~~~& 0.355~~~ & 0.0958 ~~~  &469.980 \\

			\hline \hline
		
		\end{tabular}
		\caption{{\small  Same as Table~\ref{t3}, but for  larger values of  $V_{1} $( GeV)}.}\label{t4}
	\end{center}
\end{table*}

+\end{document}